\documentclass[sigconf]{acmart}

\AtBeginDocument{%
  \providecommand\BibTeX{{%
    \normalfont B\kern-0.5em{\scshape i\kern-0.25em b}\kern-0.8em\TeX}}}

\usepackage{xspace}
\usepackage{color}
\usepackage{paralist}
\usepackage{wrapfig}
\usepackage{multirow}
\usepackage{listings}
\usepackage{makecell}
\usepackage{boxedminipage}
\usepackage{booktabs}
\usepackage{subfigure}
\usepackage{caption}
\usepackage{xcolor}
\usepackage{url}
\usepackage{tabularx}

\usepackage{breakurl} 
\usepackage[utf8]{inputenc}
\usepackage{enumitem}
\usepackage{cleveref}
\usepackage{amsthm,amsmath}
\usepackage{mathrsfs}
\usepackage{soul}
\usepackage{algorithm}
\usepackage{algorithmicx}  
\usepackage{multirow}
\usepackage{algpseudocode}  
\usepackage{graphicx}
\usepackage{threeparttable}

\copyrightyear{2021}
\acmYear{2021}
\setcopyright{acmcopyright}\acmConference[CCS '21]{Proceedings of the 2021 ACM
SIGSAC Conference on Computer and Communications Security}{November 15--19,
2021}{Virtual Event, Republic of Korea}
\acmBooktitle{Proceedings of the 2021 ACM SIGSAC Conference on Computer and
Communications Security (CCS '21), November 15--19, 2021, Virtual Event, Republic
of Korea}
\acmPrice{15.00}
\acmDOI{10.1145/3460120.3484546}
\acmISBN{978-1-4503-8454-4/21/11}

\begin{document}
\fancyhead{}

\title{Dissecting Click Fraud Autonomy in the Wild}

\author{Tong Zhu}
\email{tongzhu@sjtu.edu.cn}
\affiliation{%
  \institution{Shanghai Jiao Tong University}
}

\author{Yan Meng}
\email{yan\_meng@sjtu.edu.cn}
\affiliation{%
  \institution{Shanghai Jiao Tong University}
}

\author{Haotian Hu}
\email{hht971026@sjtu.edu.cn}
\affiliation{%
  \institution{Shanghai Jiao Tong University}
}

\author{Xiaokuan Zhang}
\email{Zhang.5840@osu.edu}
\affiliation{%
  \institution{The Ohio State University}
}

\author{Minhui Xue}
\email{jason.xue@adelaide.edu.au}
\affiliation{%
  \institution{The University of Adelaide}
}

\author{Haojin Zhu}
\authornote{Haojin Zhu (zhu-hj@sjtu.edu.cn) is the corresponding author.}
\email{zhu-hj@sjtu.edu.cn}
\affiliation{%
  \institution{Shanghai Jiao Tong University}
}

\pagestyle{empty}
\newcommand{\tong}[1]{\textcolor{cyan}{[Tong: #1]}}
\newcommand{\xz}[1]{\textcolor{blue}{[XZ: #1]}}
\newcommand{\hj}[1]{\textcolor{red}{[HJ: #1]}}
\newcommand{\tang}[1]{\textcolor{brown}{[Tang: #1]}}
\newcommand{\haojin}[1]{\textcolor{red}{[Haojin: #1]}}
\newcommand{\yan}[1]{\textcolor{brown}{[Yan: #1]}}
\newcommand{\Q}[1]{\textcolor{orange}{[Question: #1]}}

\newcommand{\bheading}[1]{{\vspace{4pt}\noindent{\textbf{#1}}}}
\newcommand{\iheading}[1]{{\vspace{4pt}\noindent{\textit{#1}}}} 

\newcommand{\etal}{\emph{et al.}\xspace}
\newcommand{\etc}{\emph{etc}\xspace}
\newcommand{\ie}{\emph{i.e.}\xspace}
\newcommand{\eg}{\emph{e.g.}\xspace}

\newcommand{\figurewidth}{\columnwidth}
\newcommand{\secref}[1]{\mbox{Sec.~\ref{#1}}\xspace}
\newcommand{\secrefs}[2]{\mbox{Sec.~\ref{#1}--\ref{#2}}\xspace}
\newcommand{\staticsecref}[1]{\mbox{Sec.~{#1}}}
\newcommand{\figref}[1]{\mbox{Fig.~\ref{#1}}\xspace}
\newcommand{\tabref}[1]{\mbox{Table~\ref{#1}}\xspace}
\newcommand{\appref}[1]{\mbox{Appendix~\ref{#1}}\xspace}
\newcommand{\eqnref}[1]{\mbox{Eqn.~\ref{#1}}\xspace}
\newcommand{\eqnsref}[2]{\mbox{Eqns.~\ref{#1}--\ref{#2}}\xspace}
\newcommand{\ignore}[1]{}

\newcommand{\sysname}{\texttt{ClickScanner}\xspace}
\newcommand{\attack}{\texttt{humanoid attack}\xspace}
\newcommand{\attacks}{\texttt{humanoid attacks}\xspace}
\newcommand{\extractor}{\texttt{Extractor}\xspace}
\newcommand{\classifier}{\texttt{Classifier}\xspace}
\newcommand{\estimator}{\texttt{Estimator}\xspace}

\begin{abstract}

Although the use of pay-per-click mechanisms stimulates the prosperity of the mobile advertisement network, fraudulent ad clicks result in huge financial losses for advertisers. Extensive studies identify click fraud according to click/traffic patterns based on dynamic analysis. However, in this study, we identify a novel click fraud, named \attack, which can circumvent existing detection schemes by generating fraudulent clicks with similar patterns to normal clicks. We implement the first tool \sysname to detect \attacks on Android apps based on static analysis and variational AutoEncoders (VAEs) with limited knowledge of fraudulent examples. We define novel features to characterize the patterns of \attacks in the apps' bytecode level. \sysname builds a data dependency graph (DDG) based on static analysis to extract these key features and form a feature vector. We then propose a classification model only trained on benign datasets to overcome the limited knowledge of \attacks.

We leverage \sysname to conduct the first large-scale measurement on app markets (\ie, 120,000 apps from Google Play and Huawei AppGallery) and reveal several unprecedented phenomena. First, even for the top-rated 20,000 apps, \sysname still identifies 157 apps as fraudulent, which shows the prevalence of \attacks. Second, it is observed that the ad SDK-based attack (\ie, the fraudulent codes are in the third-party ad SDKs) is now a dominant attack approach. Third, the manner of attack is notably different across apps of various categories and popularities. Finally, we notice there are several existing variants of the \attack. Additionally, our measurements demonstrate the proposed \sysname is accurate and time-efficient (\ie, the detection overhead is only 15.35\% of those of existing schemes).

\end{abstract}



\maketitle

\section{Introduction} \label{sec:intro}

The mobile advertisement (ad) market has grown rapidly over the past decades with the unprecedented popularity of smartphones. To motivate the app developer to embed the advertisers' ads in their apps, the pay-per-click (PPC) mechanism is widely deployed, in which the advertiser pays the developer according to the number of times the embedded ads have been clicked by users~\cite{haddadi2010fighting, ClickFraudDef}. 

However, the PPC mechanism also encounters the increasing threat of \textit{click fraud}~\cite{choempirical}. By adopting the strategy of click fraud, the unscrupulous developer generates ``fake'' ad click events that do not originate from real users to obtain extra payment from the ad network. For instance, an attacker can embed malicious code on fraudulent apps or third-party SDKs leveraged by other unsuspecting app developers to trigger the ad clicks automatically in the background without any human involvement. It is estimated that advertisers have lost 42 billion USD of ad budget globally in 2019 due to fraudulent activities committed via online, mobile, and in-app advertising~\cite{adlossreports}.

To defend against click fraud, both academia and industry have proposed a series of \textit{dynamic analysis} based approaches to distinguish fraudulent clicks from the legitimate clicks. These approaches fall into the following two categories: \textit{user-side}~\cite{caoadsherlock, binliuDECFA, FCFraud, MAdLife, AdSplit, ndss2021} and \textit{ad network-side} approaches~\cite{YouAreHowYouClick, fengdongfrauddroid, crussellMadfraud, nagarajaclicktok, DetectionontheAdvertiserSide,4595871}. (1) The user-side approaches rely on installing an additional patch or ad SDK on the user's device. The legitimacy of ad clicks is determined by checking whether the click pattern meets a certain rule.
(2) The ad network-side schemes determine whether an app performs fraudulent clicks based on statistical information (\eg, timing patterns) of the clicks through traffic analysis. 
These existing detection schemes either require users to install patches on their smartphones, which is not user-friendly, or require the ad network to collect traffic data from thousands of apps, which is less scalable. Moreover, both approaches use dynamic analysis, which is not complete since they do not cover all feasible program paths. 
Furthermore, the effectiveness of these dynamic analysis based approaches relies on the assumption that fraudulent click patterns are distinguishable from those of real users. 
Therefore, it is natural to raise the following question: \textit{Is there a smart attacker who can simulate a real human's clicks patterns and bypass existing click fraud detection?}

In this study, we answer the above question by identifying  emerging automated click fraud, named \attack. In this paper, we define \attack as a kind of click fraud that has almost the same click and traffic patterns as normal clicks. Specifically, the fraudulent applications could randomize the click coordinates/time interval, or even follow the legitimate actions of a real user to generate the clicking traffic, rendering the fake click sequences to be indistinguishable from legitimate ones even if the ad traffic is monitored. Some fraudulent applications will also receive the fake click's configuration from a remote server and avoid detection adaptively and locally. To date, the detection of \attacks via large-scale \textit{static analysis} has received little attention. Therefore it is crucial to understand and mitigate \attacks.

A large-scale static analysis of \attacks imposes the following technical challenges. 
1) How can we capture the fraudulent behavior patterns at the bytecode level by defining a set of novel features to distinguish the codes triggering false clicks from the codes generating legitimate clicks? 2) Based on the proposed features, how can we build a novel system that can automatically extract these features and accurately identify the fraudulent apps while considering very few positive samples in practice?

To address these challenges, we propose \sysname, a lightweight and effective static analysis framework to automatically detect \attacks. 
\textit{First}, our work starts from a preliminary study that aims to investigate what features can be adopted to identify \attacks. To achieve this, we build a simple prototype based on Soot~\cite{soot} to investigate the working logic behind the suspicious fraudulent apps, which likely manipulate the \texttt{MotionEvent} object to generate fake, yet indistinguishable click sequences.
\textit{Second}, through the preliminary vetting results of prototypes and careful manual checking of suspicious apps' working behaviors and bytecodes, we identify 50 apps conducting legitimate clicks and 50 apps conducting \attacks as the \textsc{seed apps} for accuracy tests and feature definition.\footnote{In order to add as many benign examples as possible to the dataset for training, we not only collect legitimate clicks on the view with the ad but also with other content. If there is no special emphasis in the latter part, the benign datasets will include legitimate clicks on views of other content. The fraudulent datasets are all fake clicks on the ad view.} Our study reveals that the \attack mainly utilizes the combination of the following four strategies to obfuscate its fake clicks and avoid detection: \textit{1) simulating the human clicks by randomizing the coordinates; 2) making the trigger condition of the fake clicks unpredictable by randomizing the triggering time; 3) generating the fake clicks by following the legitimate actions of real people; 4) predefining fake click's execution logic in code, receiving the click's coordinates and trigger condition from a remote server, and avoiding the detection adaptively and locally.}
\textit{Third}, after characterizing the working logic of \attacks, to achieve light-weight detection, we propose a novel data dependency graph (DDG) to extract key features related to the \attack. From the generated graph, a lightweight feature vector with 7 dimensions is obtained. 
\textit{Finally}, to overcome the issue of the lack of positive examples of \attacks, we exploit variational AutoEncoders (VAEs) to build a robust classifier to perform one-class classification, which flags the fraudulent apps by the reconstruction error between the input and output with limited knowledge of positive examples.

We utilize \sysname to conduct the first large-scale measurement on the \attack. The main results and contributions of our measurements are shown as follows.

\begin{itemize}[leftmargin=*]
    
    \item \textbf{Designing \sysname to dissect the \attack.}
    We identify an novel pattern of automated click fraud, named \attack, and design and implement the first tool to detect such an attack based on static analysis and VAEs with limited knowledge of fraudulent examples.
    
    \item \textbf{Effectiveness of \sysname.}
    We apply \sysname in the wild on 20,000 top-rated apps from Google Play and Huawei AppGallery to demonstrate that it can indeed scale to markets. We identify a total of 157 fraudulent apps out of the 20,000 apps with a high precision rate of 94.6\%. Some of them are popular, with billions of downloads. In terms of time overhead, the average detection time of \sysname is 18.42 seconds, which is only 15.35\% of the best case within four popular dynamic analysis based schemes (\ie, FraudDetective~\cite{ndss2021}, FraudDroid~\cite{fengdongfrauddroid}, MadFraud~\cite{crussellMadfraud}, DECAF~\cite{binliuDECFA}, and AdSherlock~\cite{caoadsherlock}). We compare the performance of \sysname with 65 existing detection engines (\eg,  Kaspersky~\cite{kaspersky}, McAfee~\cite{mcafee}) from VirusTotal~\cite{vt}. We show that 115 fraudulent apps out of the detected 157 fraudulent apps can bypass all employed engines, which demonstrates that our \sysname outperforms existing detection engines. We further apply \sysname on 100,000 apps randomly selected from Google Play.\footnote{Due to the lower number of available apps, we skip this measurement on the Huawei App Gallery.} In total, 584 apps are marked as fraudulent. We also find the difference in the behavior of \attacks between popular and niche apps as shown in Section~\ref{extensive analysis}. Overall, the experimental results demonstrate that \sysname is effective and efficient in detecting the \attack.
    
    \item \textbf{Novel findings are identified by \sysname.}
    A measurement study demonstrates the following interesting findings: 1) The \attack distribution among app categories are notably different across different app markets (\ie, Google Play and Huawei AppGallery), indicating attackers and users in different regions have different biases towards mobile ads. 2) Instead of changing the local codes of apps, the proportion of ad SDK-based attacks (\ie, the fraudulent codes are in the third-party ad SDKs) has increased from 14\% in June 2018 to 83\% in August 2020, indicating that the SDK based attack is now dominant. 3) The ad SDK-based attacks undergo a decrease after July 2020, which is possibly due to the strict security policies of app markets as shown in Section~5.3.2. 4) More sophisticated click fraud other than coordinated or timing randomization attacks is identified by \sysname, and the details are shown in Section~\ref{sec:case study}. 
\end{itemize}

\section{Preliminaries} \label{sec:background}

\subsection{Mobile Advertising Ecosystem}

\begin{figure}
  \centering
  \includegraphics[width=.95\linewidth]{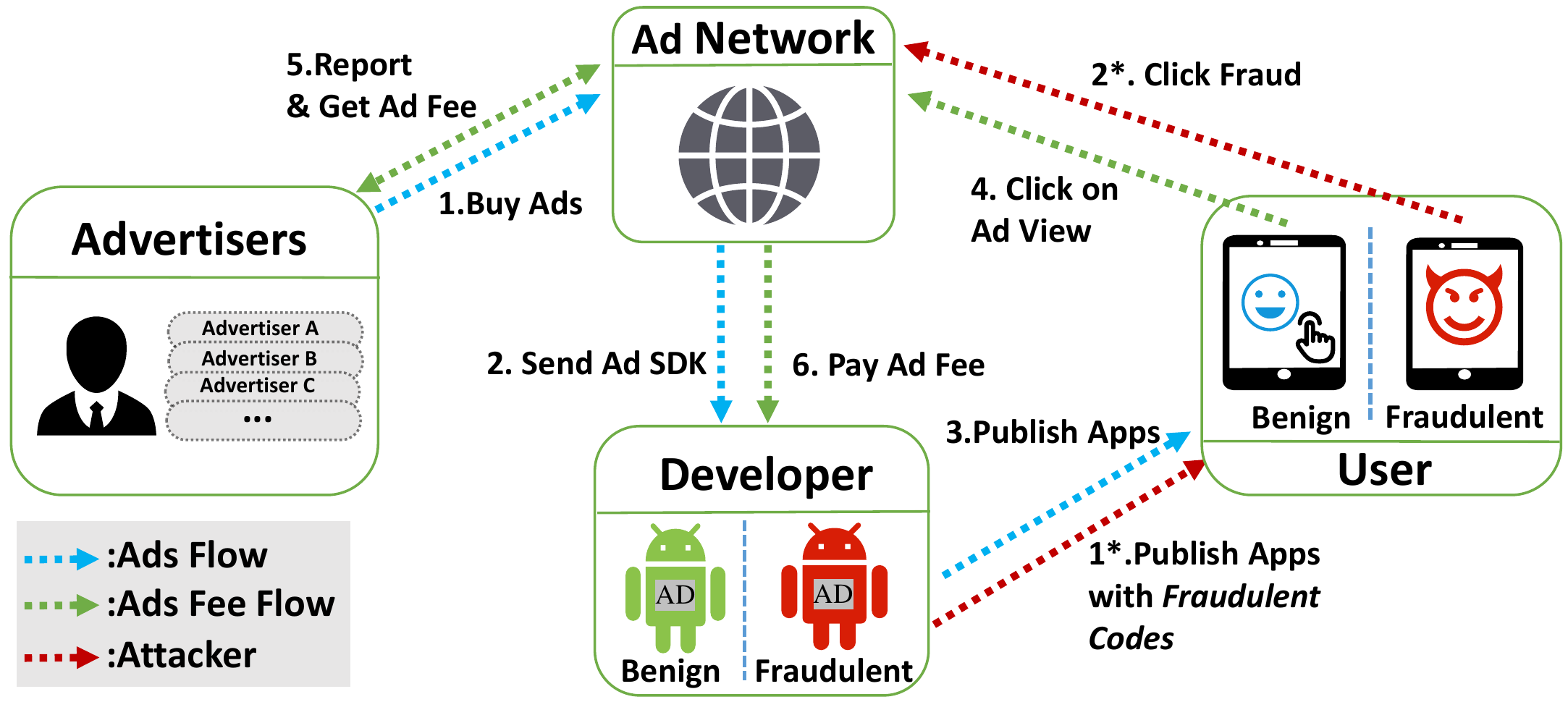}
  \caption{Overview of the mobile advertising ecosystem.}
  \label{fig:ecosystem}
\end{figure}

A typical mobile advertising ecosystem consists of four components: the advertiser, user, ad network, and developer. As shown in Fig.~\ref{fig:ecosystem}, the ad network serves as the intermediary among the other components. The advertisers publish ads in the ad network that are then embedded in the apps developed by developers. Then, developers publish apps to the users and receive the advertising fee paid by advertisers through the ad network when users click on the ads. 
Currently, one of the most popular payment mechanism in the ad network is pay-per-click (PPC), in which the revenue received by developers is related to the number of clicks.
However, these mechanisms are vulnerable to click fraud, in which attackers generate fake clicks to cheat both advertisers and users. 
For instance, researchers point out that around 10\% to 15\% of ads in Pay-Per-Click online advertising systems are not authentic traffic~\cite{adnet1,clickfraudeachyear1,clickfraudeachyear2, js1, js2}. A report published by \textit{Juniper Research}~\cite{juniper} reveals that the advertiser's loss caused by click fraud reached \$42 billion in 2019.

\begin{figure}[t]
 \begin{minipage}[t]{0.49\linewidth}
  \subfigure[The normal click's mechanism.]{
  \label{fig:click_mechanism:a}
  \includegraphics[height = 2.6cm]{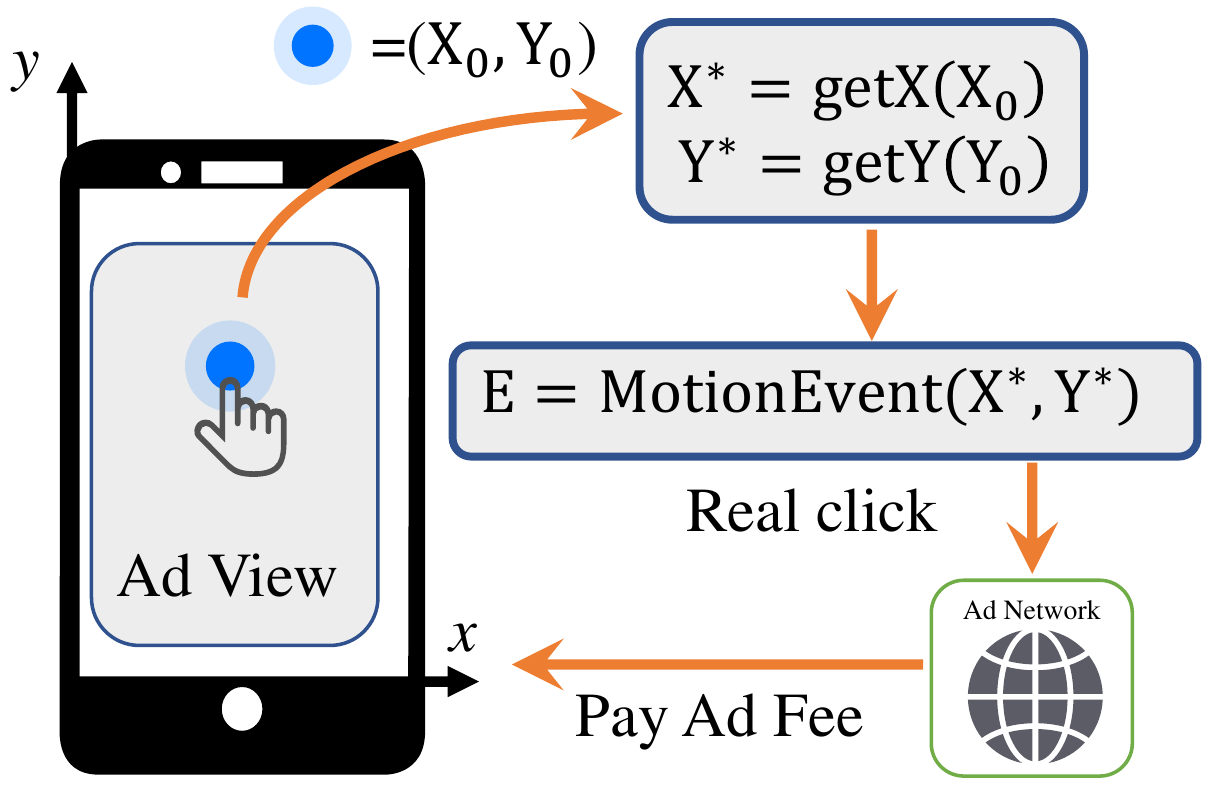}}
 \end{minipage}
 \begin{minipage}[t]{0.49\linewidth}
  \subfigure[The click fraud's mechanism.]{
  \label{fig:click_mechanism:b}
  \includegraphics[height = 2.6cm]{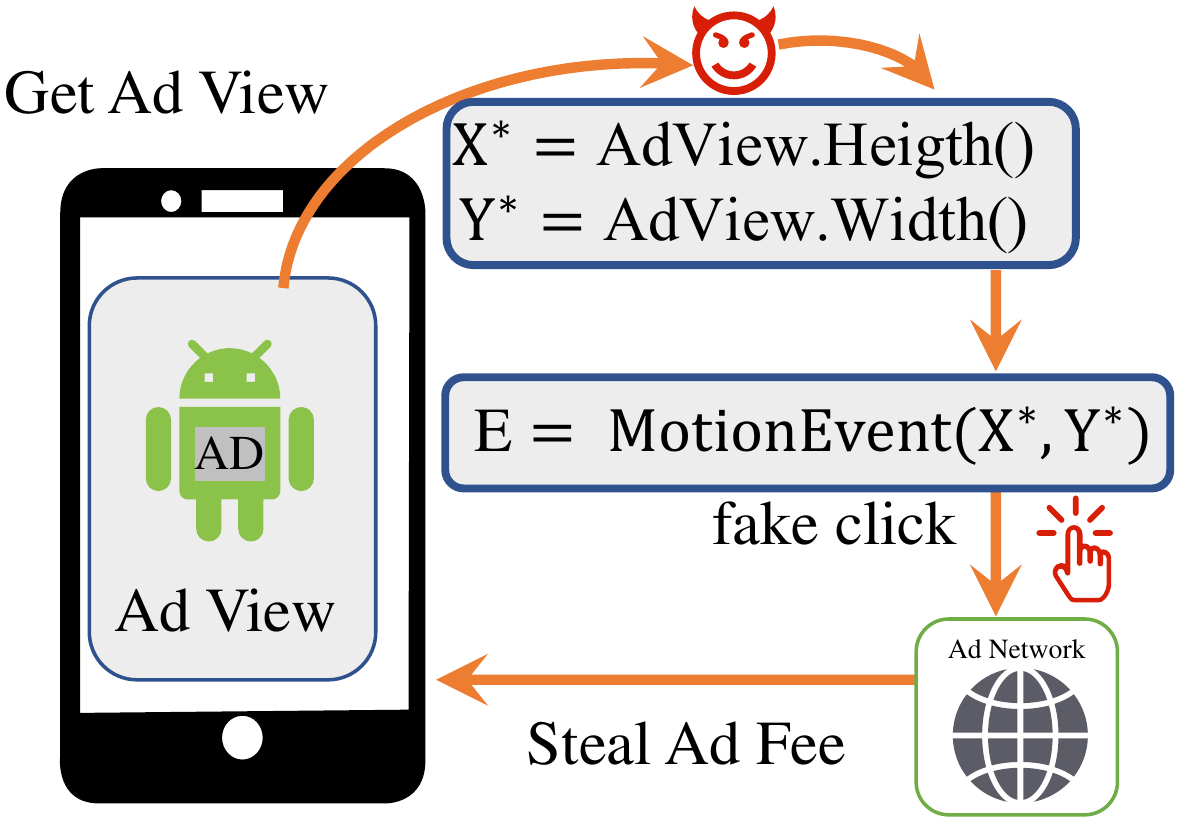}}
 \end{minipage}
    \caption{The click event generation mechanism in Android and how attackers use it to commit click fraud.}
    \label{fig:click_mechanism}
\end{figure}

\subsection{Click Event Generation Mechanisms in Android}
\label{subsec:Click Event Generation}
Since click fraud causes huge losses in the mobile ad ecosystem, it is important to figure out how click events are generated. The click mechanisms in the Android platform are shown below:

\begin{itemize}[leftmargin=*]
     \item \textbf{Normal click generation.} As shown in Fig.~\ref{fig:click_mechanism:a}, when the smart phone's screen is touched by the user, the click properties, such as time, type, and coordinates, are included in a \textit{MotionEvent} object and dispatched by the function \texttt{dispatchTouchEvent} to the targeted view. Then, the click information is delivered to the ad network, thereby an ad click is finished and counted by the ad network.
    \item \textbf{Click fraud generation.} As shown in Fig.~\ref{fig:click_mechanism:b}, in a click fraud scenario, the attacker could inject malicious code snippets into apps to generate fake clicks without any user interaction. Different from normal clicks, the attacker creates a \textit{MotionEvent} object filled with a subset of motion values (\eg, click coordinates $(X^*, Y^*)$) which are carefully fabricated. Since the \textit{MotionEvent} object can be constructed arbitrarily by the attackers, from the view of the ad network, the fake click has the same format as a normal one.
    
\end{itemize}

\subsection{Existing Click Fraud Detection Schemes}
Extensive click fraud detection schemes could be divided into two categories and their insights and limitations are shown below: 

\begin{itemize}[leftmargin=*]
    \item \textbf{User-side detection.} These schemes install an additional patch or SDK on users' devices to check the click pattern generated on users' devices.
    One of the most recent works is AdSherlock~\cite{caoadsherlock} which is based on the insight that:
    1) ``bots-driven fraudulent clicks'' can be detected because the properties are inconsistent between human clicks while remaining the same for bots-driven clicks, and 2) the ``in-app fraudulent clicks'' can be detected because the in-app clicks do not generate any motion events.
    However, there are many click fraud apps that can generate motion events that simulate the properties of a human's click through the \texttt{MotionEvent.obtain()} method~\cite{motioneventobtain}, and AdSherlock failed to consider this kind of click fraud. 
    Another recent work is FraudDetective~\cite{ndss2021}, which generates the causal relationships between user inputs and observed fraudulent activity. However, FraudDetective requires a large time overhead and cannot cover all of the app’s functionalities, which makes it difficult for FraudDetective to trigger and identify the \attack discovered in this paper. 
    
    \item \textbf{Ad network-side detection.} These schemes analyze the ad requests at the ad network server. The most recent work is Clicktok~\cite{nagarajaclicktok}, which argues that unusual click-stream traffic is often simple reuse of legitimate data traffic. Thus, they try to detect click fraud by recognizing patterns that repeat themselves in the same click-stream of ads. However, to date, a large amount of click fraud does not rely on legitimate click data streams, and attackers can also carefully construct data streams similar to the pattern of legitimate click data streams to fool detectors as shown in Section~\ref{sec:motivating}. 
\end{itemize}

Furthermore, most of the above schemes are based on dynamic analysis or traffic analysis, and therefore incur limitations. These tools cannot cover all feasible program paths, and are thus not effective and impractical to deploy in the app market. These schemes also rely on the hypothesis that the patterns generated by click fraud and real clicks are distinctly different, which may not hold true when facing \attacks. 
\section{Motivating Example and Insight}\label{sec:motivating}

\subsection{Preliminary Study on Humanoid Attack} \label{subsec:field study}

While the community struggles to properly address traditional click fraud based on dynamic and traffic analysis, deception techniques used by attackers continue to evolve. 
To characterize \attacks, we conduct a preliminary study to collect several fraudulent apps towards further building \sysname. We took a straw-man strategy that the first \attack event was spotted and discussed on a security panel inside a company. As researchers in collaboration with the company, we tried to explore more events by building up ClickScanner-Beta characterizing the app's activity to scale up the detection in the wild. Note that, when a touch event occurs, the \texttt{dispatchTouchEvent}~\cite{dispatchte} delivers the event from an Android Activity down to the target view. Therefore, we build ClickScanner-Beta based on Soot~\cite{soot} to monitor the \texttt{MotionEvent.obtain} invocation, which generates and delivers the \textit{MotionEvent} object –- an object used to report movement events~\cite{motionevent} to the \texttt{dispatchTouchEvent}. We optimized ClickScanner-Beta iteratively by first filtering out the seed apps through manual verification. When a feature is established, we update ClickScanner-Beta to reduce the overhead of manual verification.
\textit{We highlight that the difference between the \attack and random clicks is four-fold: The \textbf{basic attack} randomizes the click properties (and sometimes these properties may follow a certain distribution such as Gaussian distribution) in local codes, such as: 1) simulating the human clicks by randomizing the coordinates; 2) making the trigger condition of the fake clicks unpredictable by randomizing the triggering time. The \textbf{advanced attack} receives click properties from the cloud-server (see Section~\ref{subsec:mtxx}) or generates click properties by imposing random disturbances on user actions, such as 3) generating the fake clicks by following the legitimate actions of real people; 4) predefining fake click's execution logic in code, receiving the click's coordinates and trigger condition from a remote server, and avoiding the detection adaptively and locally.}
Through the preliminary vetting of our prototype and careful manual verification of the apps' working behaviors and decompiled codes, we identify 50 apps conducting legitimate clicks and 50 apps conducting \attacks as the \textsc{seed apps}. We elaborate on one of the most representative fraudulent apps in this section as a motivating example. 
 
\subsection{Analysis of Our Motivating Example}
\label{subsec:Analysis of motivating}

\begin{figure}
\subfigure[Visual distribution of the click points.]{
\includegraphics[width=8cm]{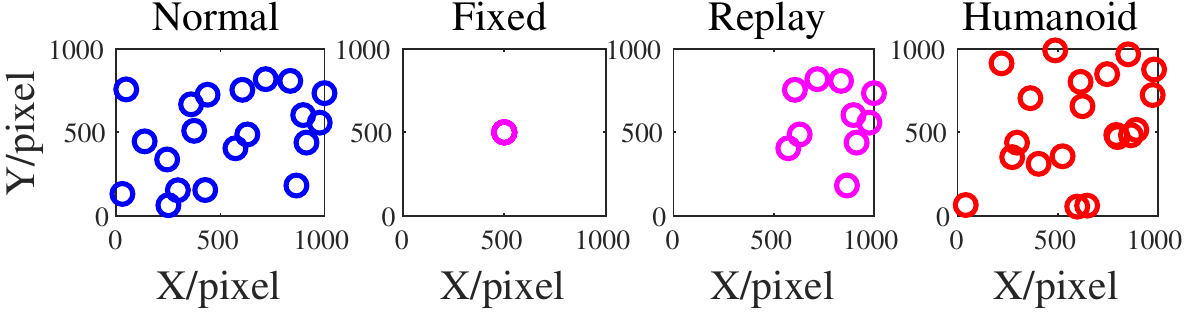}
\label{fig:motivation:coordinates}
}\vspace{-0.25cm}\\
\subfigure[Cumulative distribution of the click points along the X axes.]{
\includegraphics[width=8cm]{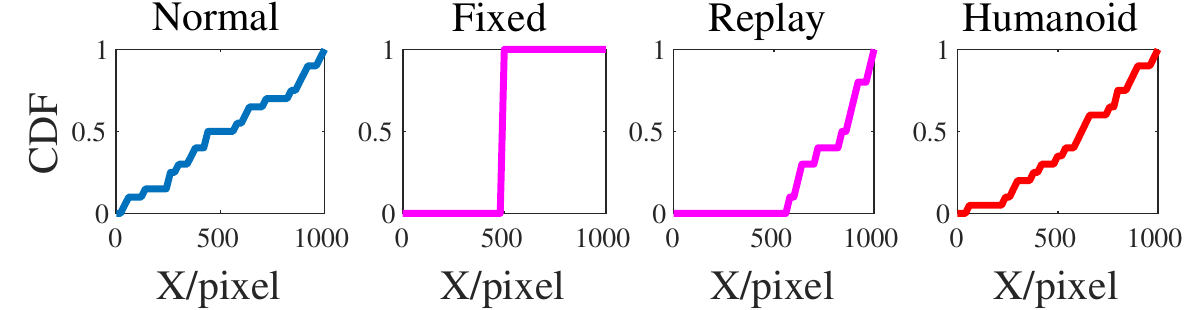}
\label{fig:motivation:cdf}
}\vspace{-0.25cm}\\
\subfigure[Time distribution of the click events.]{
\includegraphics[width=8cm]{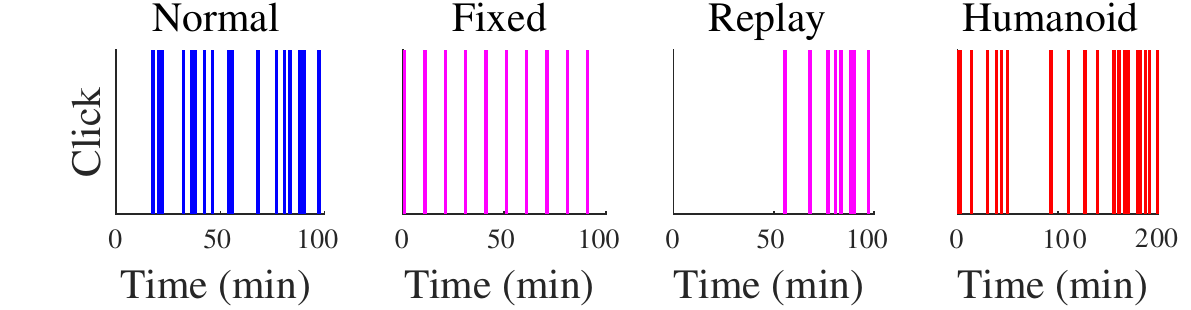}
\label{fig:motivation:time}
}
\caption{Illustration of ad clicks between normal, traditional (fixed and replay) fraud, and humanoid attack.} 
\label{fig:motivation}
\end{figure}

We use Monkey~\cite{Monkey} to randomly click on the motivating example and an app with no click fraud. We then show the coordinate distribution and timing pattern of the click event on ad banners on both apps to illustrate some of the key challenges addressed by our work. To illustrate the advantages of our work, we also reproduce traditional click frauds including \textit{fixed clicking} and \textit{replay clicking} in ~\cite{caoadsherlock, nagarajaclicktok} and compare their coordinate distributions and timing patterns with \attacks we found. The click event records are shown in Fig.~\ref{fig:motivation}. Note that for simplicity, we filter out the coordinate distributions and timing patterns generated by each kind of attack when presenting the results.

The \textit{fixed clicking} belongs to the traditional click fraud where the coordinates of the generated touch events (shown in Fig.~\ref{fig:motivation:coordinates} and ~\ref{fig:motivation:cdf}) are the same, which is easy to detect through traditional rule-based or threshold-based approaches. The \textit{replay clicking} is another traditional click fraud which replays organic clickstreams on ad banners. This can be detected by~\cite{nagarajaclicktok} because their timing patterns are similar to traditional timing patterns as shown in Fig.~\ref{fig:motivation:time}. However, we discovered that the \attack is 
more sophisticated and cannot be easily detected using the above approaches, since attackers simulate human clicks to camouflage  their false clicks. The coordinate distribution and timing pattern of the \attack is generated as if the user clicks. For instance, the distribution of coordinates in \textit{X} axis of \attack resembles normal clicking, whereas the ad traffic generated by the fraudulent app is nearly 0.5 times more than that of the normal app, which can be easily passed off as traffic from legitimate users that are interested in the ad.

\begin{figure}[b]
  \centering
  \includegraphics[width=1\linewidth]{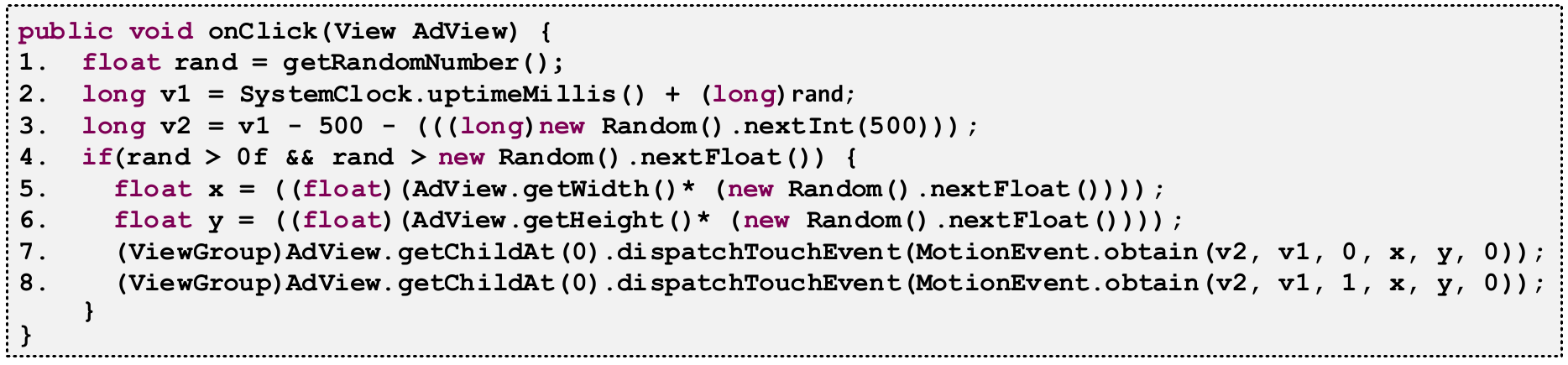}
  \caption{The code snippet simplified from the fraudulent app in the motivating example.}
  \label{fig:me_onClick_code}
\end{figure}

\begin{figure*}[t]
  \centering
  \includegraphics[width=0.9\linewidth]{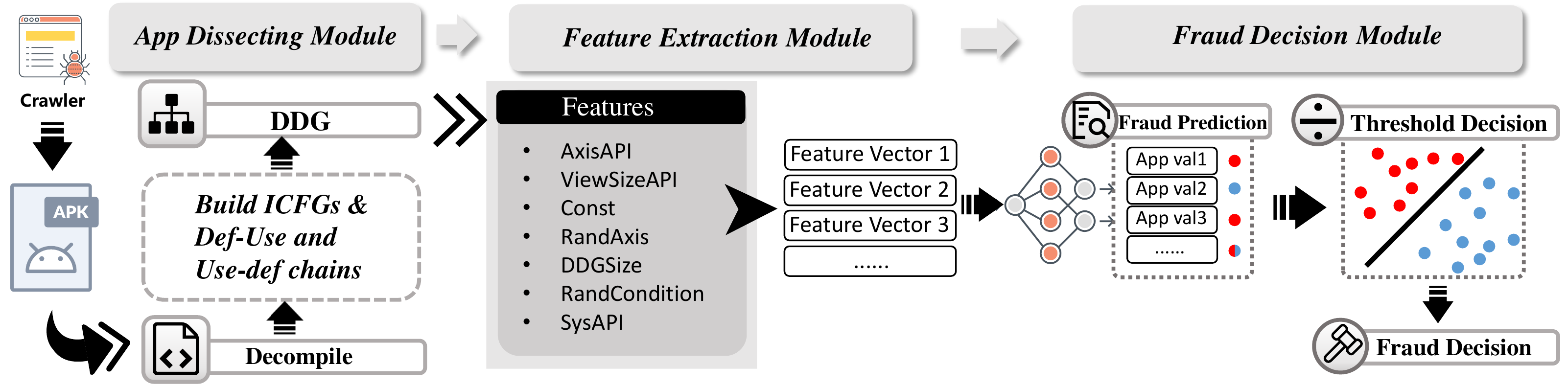}
  \caption{The workflow of \sysname.}
  \label{Workflow of ABC}
\end{figure*}

To reveal how the attacker achieves this, we next analyze the decompiled codes of the above \attack app. 
As shown in Fig.~\ref{fig:me_onClick_code}, the code snippet example simplified from the motivating example exhibits the click fraud following legitimate human actions. In general, the ad view in the fraudulent app is randomly clicked again in a random time period after the real person clicks on the ad. The fake clicks will never be triggered if the real user does not click the ad. To do this, attackers insert the function \texttt{dispatchTouchEvent} at lines 5 and 6 which generates the fake clicks in the body of the function \texttt{onClick()}. At lines 1 and 2 in Fig.~\ref{fig:me_onClick_code}, the attacker also tries to fool  detectors by making the trigger condition of the click event unpredictable and by randomizing the coordinate and trigger time of the fake click, impersonating a human's click pattern. 

This makes it hard for traditional ad network-side fraud detection approaches to detect it because the click in \attack 
is very likely to be triggered by a real person due to the uncertainty of click patterns. 
Additionally, user-side approaches are ineffective in detecting \attacks, because the click patterns and click effects on the user-side generated by \attack are almost the same as those of real clicks since the attackers are assumed to be allowed to arbitrarily construct the \textit{MotionEvent} object instead of just using the same \textit{MotionEvent}~\cite{caoadsherlock}.  Further, it is also a challenge for detectors to trigger \attacks due to its random trigger condition. Therefore, there is a pressing need to address the fake clicks that stem from the \attacks. To this end, this paper presents \sysname, a scalable, efficient, and automated static analysis system to identify the \attack.

\subsection{Insight of \sysname}

Section~\ref{subsec:Analysis of motivating} demonstrates that the \attack can manipulate ad clicks with similar pattern to that of a normal click scenario, thus causing existing detection schemes based on dynamic analysis futile. In this study, to successfully detect the \attack, the key insight is that although the click pattern is camouflaged as legitimate, at the bytecode code level the difference of the ad click trigger condition and generation process between legitimate and fraudulent apps are notably  significant, which can be characterized as detection features.

As illustrated in Fig.~\ref{fig:me_onClick_code}, it is observed that when generating a fraudulent click, the app must exploit methods \texttt{getHeight} and \texttt{getWidth} to obtain the height and width of the ad view. Furthermore, the click trigger condition defined by the method \texttt{Random} is utilized to disguise fraudulent clicks. By checking the \textit{parameters} and \textit{trigger conditions} within the bytecode of \texttt{MotionEvent}, it is feasible to detect this \attack case.
Therefore, we propose a static analysis based detection scheme \sysname, 
and break down \sysname in Section~\ref{sec:ABC}.

\section{System Design of ClickScanner}
\label{sec:ABC}

As shown in Fig.~\ref{Workflow of ABC}, \sysname mainly consists of three components (\ie, App Dissecting Module, Feature Extraction Module, Fraud Decision Module) to automatically detect \attacks. 
In this section, we break down \sysname into each component.

\subsection{App Dissecting Module}

For a given APK, \sysname first determines whether it is associated with mobile ads, and then converts the properties of the click event targeted at an ad view to a data dependency graph (DDG) for further feature extractions. 

\subsubsection{Preprocessing of App Dissecting}
When detecting \attacks in apps from the app market, it is crucial for \sysname to only focus on apps involving mobile ads. To achieve this goal, \sysname has the following three steps. First, \sysname checks apps' permissions and filters out those  with no permissions such as INTERNET and ACCESS\_NETWORK\_STATE~\cite{exclueadlibrary}. Second, for the remaining apps, \sysname leverages LibRadar~\cite{LibRadar}, a popular and obfuscation-resilient tool to detect third-party libraries on those apps and discards the apps without ad libraries. Third, \sysname needs to remove views that do not contain ad contents to avoid unnecessary analysis. Since there are no explicit labels that would allow us to easily distinguish ad views from other views, in this study, \sysname uses the relevant ad features, such as string, type, and placement features, to determine ad views, followed by prior research~\cite{fengdongfrauddroid}. 
In summary, only the apps that successfully pass the above three analysis steps would undergo the static analysis of \sysname.

\subsubsection{Extracting Click Event Properties through Static Analysis}
After \sysname selects those click events targeted at ad views, \sysname performs static analysis on them and extracts their properties and trigger conditions. As mentioned in Section~\ref{subsec:Click Event Generation}, attackers typically use the \texttt{MotionEvent.obtain} function to create a new \textit{MotionEvent} object by obtaining the properties of a click event as its parameters, and then attackers deliver it to the \texttt{dispatchTouchEvent} function to perform the \attack.
Therefore, for a given app (APK), \sysname first utilizes the static analysis tools Soot~\cite{soot} and Flowdroid~\cite{flowdroid} to build inter-procedural control flow graphs (ICFGs), Def-Use (DU), and Use-Def (UD) chains of it.
However, separately deploying the above ICFG, UD, and DU chains cannot represent the parameters assignment process of the \texttt{MotionEvent.obtain} function and the trigger condition formation process of the \texttt{dispatchTouchEvent}.
Therefore, to overcome these issues, we propose a novel data dependency graph (DDG) to show the overall \textit{properties} and \textit{trigger conditions} of the click event for further feature extraction, and the details of DDG building are introduced as follows.

\begin{figure}[t]
  \centering
  \includegraphics[width=1\linewidth]{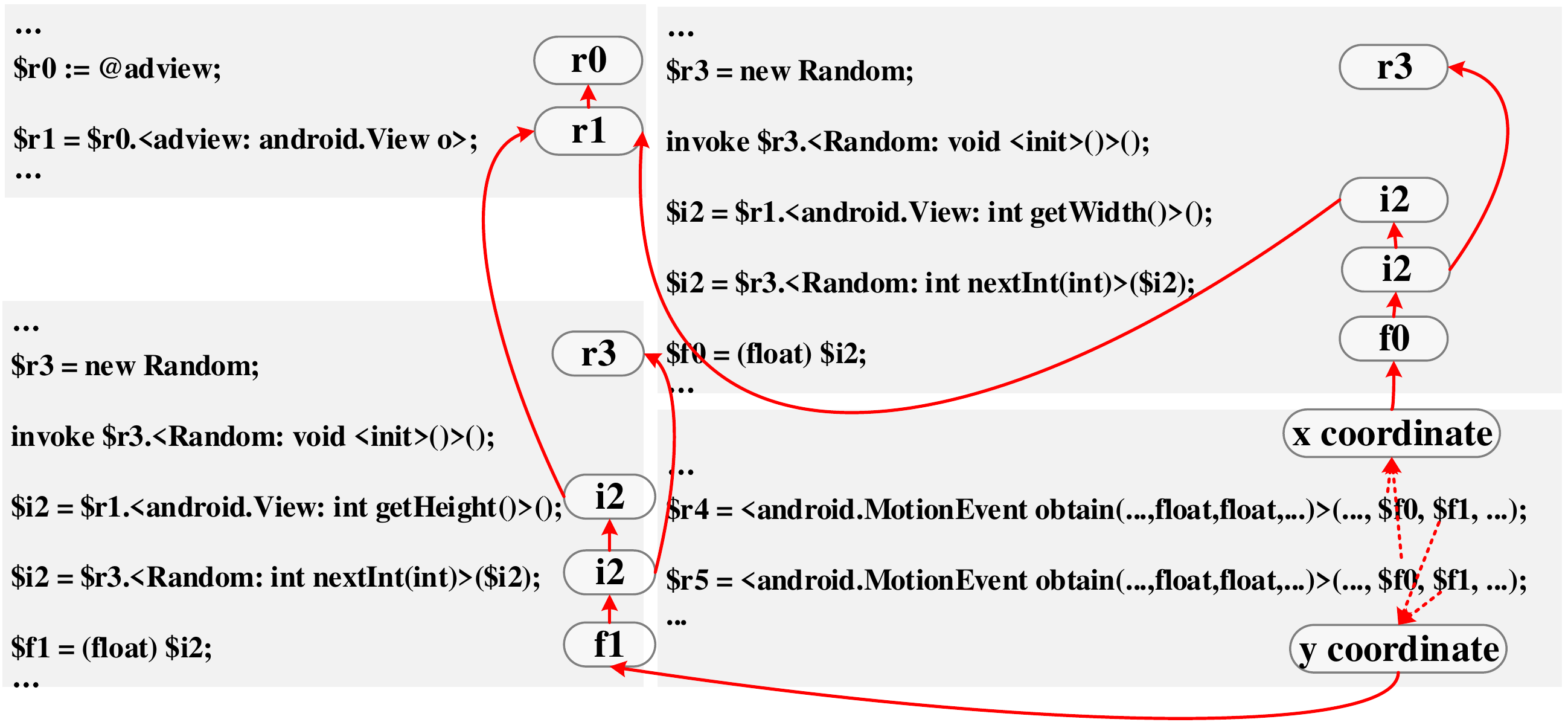}
  \caption{A simplified illustrative figure of the data dependence graph (DDG) of the motivating example constructed by \sysname for the code in Fig.~\ref{fig:me_onClick_code}. The red arrows are the routes of backward program slicing.}
  \label{fig:meddg}
\end{figure}

\noindent \textbf{The initialization of DDG.}
We propose a novel data dependency graph (DDG) to show the overall properties and trigger conditions of the click event based on ICFGs, DU, and UD chains for further feature extraction. DDG can include all the data that make up properties and trigger conditions of the click event in a graph, where each node represents the statement, and each edge represents the dependency relation between the two statements. We can find out what data have been used to form the properties and trigger conditions of the click event and what the relationship is between them by the DDG. After obtaining the ICFG, DU chains, and UD chains, we develop backward program slicing in Algorithm~\ref{alg:ddgtool} to build the DDG. Fig.~\ref{fig:meddg} shows a DDG generated by \sysname corresponding to Fig.~\ref{fig:me_onClick_code}. The red arrows are the routes of backward program slicing and nodes are the statements.
The inputs of the algorithm are the ICFG, DU and UD chains, and the \textit{root} which are those items in the condition expression of \texttt{dispatchTouchEvent} and the parameters of \texttt{MotionEvent.obtain} as mentioned in Section~\ref{sec:motivating}. In particular, the roots of this DDG in Fig.~\ref{fig:meddg} are ``x coordinate'' node and ``y coordinate'' node. The algorithm starts with the empty set \textit{data dependency graph (DDG)} and aims at finding the assignment process of those items and parameters. It is observed that both the values of the items in the condition expressions and the parameters of \texttt{MotionEvent.obtain} representing the properties of a click event are usually composed of four types of data. They are \textit{constants}, \textit{variables}, \textit{return value} of a method and \textit{parameters} of the function which calls the \texttt{dispatchTouchEvent}. One or more of these four types of data are combined through arithmetic operations to form the final result.

\begin{algorithm}[t] 
\caption{DDGTool} 
\label{alg:ddgtool} 
\begin{algorithmic}[1]
\Require ICFG; UD chain; DU chain; root;
\Ensure DDG; 
\State {DDG = emptyset}
\State {DDG.setRoot($root$)}
    \While{DDG is changing}{ 
        \For{every i in DDG}{ 
            \If {i is const}
                \State {const\_def = getDefSite($i,UD Chain$)}
                \State {DDG += New DDGNode($const\_val$)}
            \ElsIf {i is func}
                \If {!isSysAPI($i$)}
                    \State func\_def = getDefSite($ICFGs, i$)
                    \State DDG += getSubGraph($func\_def$)
                \Else 
                    \State DDG += New DDGNode($i$)
                \EndIf
            \ElsIf {i is var}
                \State DDG += New DDGNode($i$)
                \State DDG += getPre($ICFG,UD \& DU Chain,i$)
            \Else
                \State para\_caller = getCaller($ICFGs,i$)
                \State DDG += getPara($para\_caller$)
            \EndIf
        }
        \EndFor
    }
    \EndWhile
    \\
    \Return {DDG}
\end{algorithmic}
\end{algorithm}

\noindent \textbf{The expansion of DDG.}
For each item in the DDG generation, as shown at line 3 to line 22 in Algorithm~\ref{alg:ddgtool}, \sysname handles the above four types of data by repeating the following steps. 
    1) If it is a constant, \sysname will find the definition site of it by UD chain and directly add it to the DDG. 
    2) If it is a return value of a certain system method that usually has a fixed meaning, \sysname directly adds the return values of it to the DDG. 
    However, if it is a return value of the developer-defined method without fixed meaning, \sysname finds the method's definition site and identifies what processing has been performed in its method body. Then, \sysname converts all the nodes in their method bodies into subgraphs of the DDG. 
    3) If it is a variable that is usually formed by different types of data, to figure out the variable's meaning, \sysname needs to find the variable's assignment process. Therefore, \sysname finds its predecessors from the definition site of the variable based on the ICFG and UD chains and adds it to the DDG.
    4) If it is a parameter of the method in which the \texttt{dispatchTouchEvent} is called, \sysname first gets the caller of the method based on ICFG and then finds the parameter values of the method.

As shown in Fig.~\ref{fig:meddg}, \sysname successfully finds exactly where the \attack occurs in codes. The process of generating the DDG of the trigger condition is similar to the above description, so we omit it due to space limitations.

\subsection{Feature Extraction Module}\label{sec:Feature Extraction}
Once a DDG is built, \sysname can find all necessary features in it for verifying whether the \attack takes place in the given app. Android systems typically have 7 different constructors for the \texttt{MotionEvent.obtain} and all of them have the following common parameters: \textit{eventTime, actionType, axisValue and metaState}, among which we choose the axis values (\ie, touch position (AXIS\_X, AXIS\_Y)) for further study. Because the benign app receives the \texttt{MotionEvent} object from the system, sometimes it needs to record the coordinates of the click, and then dispatches it. However, for fraudulent apps (\ie, Fig.~\ref{fig:me_onClick_code} at lines 5 and 6), they first obtain the height and width of the ad view and construct the fake click's coordinates that follow a random distribution, which mimics the benign app behaviors. As a result, even if the traditional click fraud detection approaches can obtain click traffic, they cannot distinguish between a \attack and a normal click since properties such as the coordinates are similar.

This shows that instead of analyzing the pattern of generated coordinates, the \attack can be identified by checking the process of coordinate generating. For instance, the ``illegal operations'' including obtaining height/width and exploiting \texttt{Random()} method in Fig.~\ref{fig:me_onClick_code} may be used to detect its fraudulent behaviors. In a real-world scenario, we further characterize the axis into five features:

\noindent \textbf{(1) The number of APIs for getting the actual click coordinates generated by users (AxisAPI)}. As shown in Fig.~\ref{fig:me_onClick_code}, the fraudulent app involves no APIs to get the coordinates of real users' click (\eg, getX() and getY()). Instead, it constructs the coordinates by itself. Intuitively, the existence of APIs, which are used to get the actual click coordinates generated by users, can be indicative of whether the app is a fraudulent application. When an app contains a \textit{MotionEvent} whose coordinate parameters do not involve the system APIs above, we take it as a potentially fraudulent application. 
As shown in Table~\ref{F-score of single feature.}, the F-score of AxisAPI is 0.81 when identifying \attack instances over our ground truth dataset (\textsc{seed apps}).
    
 \noindent \textbf{(2) The number of APIs for getting the size of ad view (ViewSizeAPI)}. Many fraudulent apps obtain the size of the ad view in order to place the coordinates of the fake click inside the ad view. Although some benign apps will also get the view size, the proportion is much lower than that in fraudulent apps based on the observation of benign samples in our dataset.

\noindent \textbf{(3) The number of the constants (Const)}. Since some fraudulent apps try to click on the area around a fixed point in the ad view, such as the download and install button, they will obtain the size of a view and calculate it with a constant to get a specific point coordinate. The F-score for this feature is shown in Table~\ref{F-score of single feature.}.

\noindent \textbf{(4) The number of API for getting random numbers (RandAxis)}. To better mimic human clicks, the fraudulent apps often use APIs that generate random numbers (\eg, \texttt{random.nextGaussian}) when constructing click coordinates or the time distribution of clicks, as shown in the motivating example. In doing so, attackers can disguise the traffic generated by fake clicks as traffic generated by real people and make the fake clicks unpredictable to evade the dynamic analysis. Therefore, we use this as an indicator to identify fraudulent apps. The F-score for this feature, as measured on our ground-truth set, is illustrated in Table~\ref{F-score of single feature.}.

\noindent \textbf{(5) Size of the DDG (DDGSize)}. Our manually labeled dataset shows that fraudulent apps tend to process the data several times before passing it to the \texttt{MotionEvent.obtain} as its coordinate parameters, while benign apps tend to directly take the return value of the methods like \texttt{getX} as the coordinate parameters. The larger size of the DDG indicates that the data have been processed more times before being passed to the \texttt{MotionEvent.obtain}. The F-score for this feature is shown in Table~\ref{F-score of single feature.}.

Meanwhile, the attackers also tend to change their behaviors to evade detection, which can be detected by analyzing the trigger conditions of the click events. The unique software and hardware resources on mobile devices enable fraudulent apps to cover their behaviors with a wider spectrum of triggers, that is, conditions under which the hidden operations will be performed~\cite{HSO}. For example, in Fig.~\ref{fig:me_onClick_code} at line 4, the fraudulent app tries to fool detectors by randomizing the trigger condition of the click event, which is like a human's click timing pattern and difficult to be triggered by dynamic analysis. Therefore, we also focus on the trigger condition of click events and characterize it into two features:

\noindent \textbf{(6) Random Numbers in Condition Expression (RandCondition).}
Many fraudulent apps tend to randomize the trigger conditions and trigger frequency of \attacks to simulate legitimate clicks, which makes the fake clicks indistinguishable and undetectable. Additionally, dynamic analysis requires much time to interact with these apps so it is difficult to cover all paths of the \attack. Hence, we regard the invocations of functions in the process formation of the trigger conditions, which can generate random numbers as a feature to identify the \attack.

\noindent \textbf{(7) System Call in Condition Expression (SysAPI).}
Some hidden sensitive instances with a similar purpose to the \attack have been discussed~\cite{HSO, Triggerscope}. They are subject to some system properties or environment parameters (\ie, OS or hardware traces of a mobile device). They can only be exposed to an app through system interfaces. Hence, we can infer that the condition of the \attack is also expected to involve, directly or indirectly, one or more API calls for interacting with the OS, and we regard them as another feature. The F-score for this feature is 0.61 as illustrated in Table~\ref{F-score of single feature.}.

\begin{table}[t]
\centering
\caption{F-score of features}
\begin{threeparttable}
\begin{tabular}{llll}
\hline
                 \textbf{AxisAPI}         & \textbf{ViewSizeAPI}     & \textbf{Const}           & \textbf{RandAxis}     \\ \hline  
                  \multicolumn{1}{c}{0.81} & \multicolumn{1}{c}{0.79} & \multicolumn{1}{c}{0.83} & \multicolumn{1}{c}{0.62} \\ \hline
                 
                  \textbf{DDGSize}         & \textbf{RandCondition}   & \textbf{SysAPI}     &     \\ \hline
  \multicolumn{1}{c}{0.82} & \multicolumn{1}{c}{0.54} & \multicolumn{1}{c}{0.61} & \\ \hline
\end{tabular}
\begin{tablenotes}    
\footnotesize               
\item[1] F-score is calculated based on classification with each single feature.          
\end{tablenotes}           
\end{threeparttable}
\label{F-score of single feature.}
\end{table}

Although all the above features can contribute to the detection of \attacks to a certain degree, certain kinds of \attacks may involve several features and a single feature may cause high false positives and negatives. Therefore, none of those features can work alone.~\cite{zheng2018smoke} Hence, our key idea is to use some of these features collectively. We finally combine all 7 features into the same feature space according to our experiment result, which is illustrated in Section~\ref{Detection Accuracy}.
Furthermore, we apply normalization to the features before feature vectors formalization because the components of the features are different. To determine the weight of each feature, the entropy weight method is deployed by \sysname.

\subsection{Fraud Decision Module}\label{subsec:classifier}

Existing click fraud detection models either need to specify many rules for classification~\cite{nagarajaclicktok, caoadsherlock, fengdongfrauddroid, binliuDECFA}, which leads to high false negatives due to the incomplete and statistically unrepresentative rules, or require a large number of malicious samples as the training set~\cite{crussellMadfraud}, which is unrealistic due to the lack of labeled datasets.  
Moreover, since these existing approaches rely heavily on the knowledge of certain rules and training set labels, they may fail to handle subsequent variant click fraud. 
To overcome these limitations, we build an effective classifier based on Variational AutoEncoders (VAEs) with limited knowledge about fraudulent examples. This can reduce the researchers' dependence on fraudulent data sets and is more robust to variants of such newly discovered attacks.

In a nutshell, a VAE is an autoencoder whose encoding distribution is regularized during training in order to ensure that its latent space has good properties so that it can be used to generate new data that is similar to the inputs. We use benign examples to train our classifier and determine whether an input is benign or not according to the reconstruction error between the input and output. Specifically, the encoder is a neural network. Its input is $x$, which is the feature vectors generated by the \sysname. The encoder's output is a hidden representation $z$, which is the aforementioned latent space. The encoder will perform dimensionality reduction on the input $x$ because the encoder must learn an efficient compression of the data into this lower-dimensional space. The decoder is another neural network. Its input is the representation $z$, and its outputs are the parameters to the probability distribution of the data with weights and biases $\phi$. 
Some information may be lost due to the dimensionality reduction of the encoder, and some new data are generated due to the random sampling of the decoder. We can use the reconstruction error to measure the difference between the input and output.

In the training phase of our classifier, it is trained with benign examples' feature vectors in advance so that its encoder will be able to learn the representations of benign examples. To do this, we randomly selected 10,000 benign apps from~\cite{androzoo} for training. We train the VAE with feature vectors of those APKs that are not marked by all the engines from VirusTotal~\cite{vt}, a website that aggregates many antivirus products and online scan engines to check for viruses. Although, as mentioned in~\cite{virustotalmeasurement}, the detection results of VirusTotal are not always reliable, because we use many benign samples for training, a relatively small number of fraudulent samples that are not detected by VirusTotal will not affect the distribution of the latent space. Once the classifier has been well trained, its encoder will learn benign examples' representations in the latent space. 
After training, we feed a tested app's feature vector, which is extracted from the newly formed DDG in the \extractor, to 
the VAE and output the reconstructed feature vector containing the information of the latent space in the training phase. Our classifier will consider its input to be fraudulent only when its reconstruction error exceeds a certain threshold $t$. It is similar to building a borderline that encompasses all benign examples so that we only need to check whether an input is in the borderline by computing the reconstruction error. 

To the best of our knowledge, our model is the first Android \attack detector with limited knowledge about fraudulent examples. Due to the lack of malicious examples in reality, this makes \sysname practical to deploy.
\section{Measurements} \label{sec:evaluaiton}

As mentioned in Section~\ref{subsec:field study}, due to the absence of existing benchmarks in this research area, we manually label 100 apps containing 50 fraudulent examples and 50 benign examples as our \textsc{seed apps} for fine-tuning and accuracy tests. Then we utilize \sysname to conduct the first large-scale measurement 
of the \attack in the current app market based on 120,000 apps (10,000 top-rated apps from Google Play, 10,000 top-rated apps from Huawei AppGallery, and 100,000 randomly selected apps from Google Play), and elaborate on several important findings. All experiments are performed on a Windows 10 Desktop, equipped with 8 CPU Cores at 3.6GHz and 32 GB of RAM.

\subsection{Evaluation of ClickScanner} \label{Detection Accuracy}

\begin{table}
\centering
\caption{Performance of the classifier}
\label{Detailed performance of classifier}
\begin{tabular}{cccc}
\hline
                           & \textbf{Precision} & \textbf{Recall} & \textbf{F-score} \\ \hline
\textsc{Seed Apps} & 48/50 = 96\%                                  & 48/50=96\%                               & 0.960                                 \\ \hline
\end{tabular}
\end{table}

\subsubsection{Fine-tuning of \sysname}
\begin{figure}
\centering
\includegraphics[height=5.5cm]{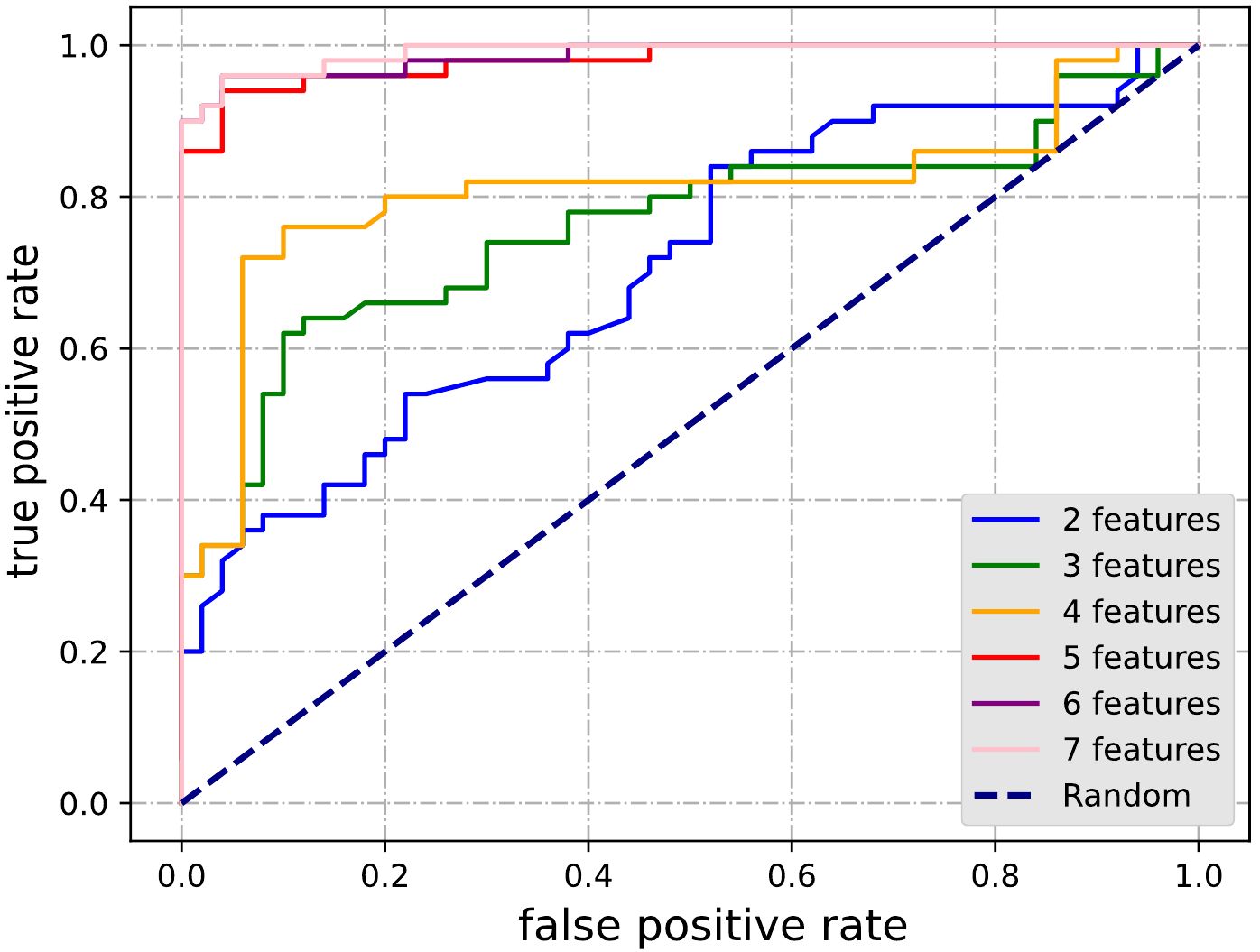}
\caption{The ROC curve for different combination of features.} 
\label{fig:ROC}
\end{figure}

Before utilizing \sysname to conduct large-scale analysis, it is necessary to fine-tune the \sysname's parameters on \textsc{seed apps} to achieve the best performance. As mentioned in Section~\ref{sec:Feature Extraction}, there are seven different features for \sysname. To determine the best feature combinations, we traverse all combinations from 2 to 7 features and show their best performances with the ROC (Receiver Operating Characteristic) curves in Fig.~\ref{fig:ROC}. It is observed that the performance is improved by adding new features. 
When there are 5 features selected ($AxisAPI$, $ViewSizeAPI$, $RandAxis$, $DDGSize$, $RandCondition$), adding more features only leads to a slight improvement 
in accuracy, which demonstrates that our feature vector can adequately describe the app behaviors and help classifiers to identify fraudulent apps. 
We also evaluate \sysname under different thresholds $t$ for threshold selection. It is observed when $t$ is set to 2.04, \sysname achieves the best performance. 
In the following measurement study, to achieve the best accuracy, $t$ is set to 2.04 and all 7 features are selected by \sysname. 

\subsubsection{Effectiveness of \sysname}
Table~\ref{Detailed performance of classifier} shows when choosing the above parameters, 48 apps out of 50 fraud apps in the \textsc{seed apps} are successfully recognized by \sysname. \sysname achieves the F-score of 0.960, showing its effectiveness in detecting \attacks. 
There are 2 false positive and 2 false negative cases and we discuss the root causes in Section~\ref{sec:discussion}. 
Note that since \textsc{seed apps} are independently extracted from manual check (see Section~\ref{subsec:field study}), \sysname achieves high average values of precision and F1-score with limited knowledge about malicious examples, which implies its effectiveness. The effectiveness is also substantiated with the high precision of detection in the wild (see Section~5.2.2).

\subsubsection{Comparison with the State-of-the-Art Ad Fraud Detection Tool} \label{ssssec:Comparison with FraudDetective}
Currently, the most up-to-date fraud detection tool is FraudDetective implemented by Kim et al.~\cite{ndss2021}, which computes a full stack trace from an observed ad fraud activity to a user event by connecting fragmented multiple stack traces. It is an effective tool which could detect three types of ad fraud. However, like other tools that use dynamic analysis, it incurs a large time overhead to execute apps and interact with them. Additionally, since some fraudulent apps will randomly trigger the \attack, FraudDetective may not be able to cover all the program paths, and thus it is difficult for FraudDetective to trigger all \attacks discovered in our study. 

Since FraudDetective and their datasets are not publicly available, we make our best effort to craft the datasets of ClickScanner as similar as possible to the ones used by FraudDetective, and we acknowledge that data duplication may exist between ClickScanner and FraudDetective. The two datasets were obtained around the same time with a similar data collection methodology. In our \textsc{basic dataset}, we collected 10,000 top-rated apps in total from each category from Google Play updated in July 2020, and we also conducted a longitudinal study, collecting fraudulent apps with different versions published from August 2017 to December 2020. The researchers of FraudDetective collected the top 10,024 apps from each of the Google Play categories from April 2019 to September 2020 and randomly sampled additional 38,148 apps from APK mirror sites. We believe that these two datasets are likely to have overlapped apps. However, FraudDetective may fail to trigger humanoid attacks discovered in this paper since it is based on dynamic analysis alone. 
In their evaluation part, FraudDetective did not detect any app that generates a forged click among the 48,172 apps crawled from the Google Play Store. By contrast, \sysname successfully identified 157 fraudulent apps among 20,000 apps in the \textsc{basic dataset} as shown in Section 5.2.1. In summary, \sysname outperforms existing detection tools in the aspect of detecting \attacks, and we leave the comparison on the same large-scale dataset for future work. 

\subsection{Detecting Humanoid Attacks in App Markets}
We first build a \textsc{basic dataset} with 20,000 top-rated apps and show the detection results of \sysname below. We show more details of the fraudulent apps we have detected in Appendix ~\ref{appen:Results of Fraudulent Apps and SDKs}.

\subsubsection{The Collection of the \textsc{Basic Dataset}}
The \textsc{basic dataset} contains 20,000 apps, in which every 10,000 app is top-rated and sorted by downloading numbers in each category from Google Play and Huawei AppGallery updated in July 2020. We choose these apps because, in app markets, the few most popular apps contribute to a high majority of app downloads. Therefore, by analyzing the \textsc{basic dataset}, we can focus on \attack cases with the highest influence~\cite{App_downloads}. 
Additionally, Google Play and Huawei AppGallery are the biggest app markets in the U.S. and China respectively, which ensure that our study is representative in scope. \textit{In this study, if not otherwise specified, all analysis is done using the \textsc{basic dataset}.}

\subsubsection{The Humanoid Attack Cases Detected by \sysname}\label{sec:homanoid:attack}
After conducting analysis of 20,000 apps, \sysname identifies 170 suspicious click activities from 166 suspicious apps. Then, by manually checking those apps' decompiled codes, 74 \attack activities in 63 apps and 140 \attack activities in 94 apps from Google Play and Huawei AppGallery are found respectively. The precision rate of 94.6\%  over 20,000 apps demonstrates the effectiveness of our classifier and fine-tuned parameters. 
As a consequence of over 1.2 billion downloads of such fraudulent apps in the market, the \attack is very likely to have deceived both advertisers and users with fake ad clicks which corresponds advertisers huge losses. We attempted to perceive the status quo (until March 29, 2021) of 63 fraudulent apps and 94 fraudulent apps in the \textsc{basic dataset} we detected from Google Play and Huawei, respectively. We found that 13 of 63 in Google and 26 of 94 in Huawei have been removed, but the remaining apps are still publicly available. We are also in the process of liaising with relevant app vendors for responsible disclosure. And Google notified us that they have received our report. 

To quantify the damage of \attacks in the real world, it is important to know the category distributions of identified malicious apps. Thus, for each fraudulent app, we extract its category (\eg, books, education, weather) labeled by app markets. And analyze the distribution of Apps affected by \attacks and show them in the Appendix ~\ref{appen:humanoid:category} due to the page limitation.

\subsubsection{Comparison with Existing Detection Engines}

\begin{figure}
  \centering
  \includegraphics[width=1\linewidth]{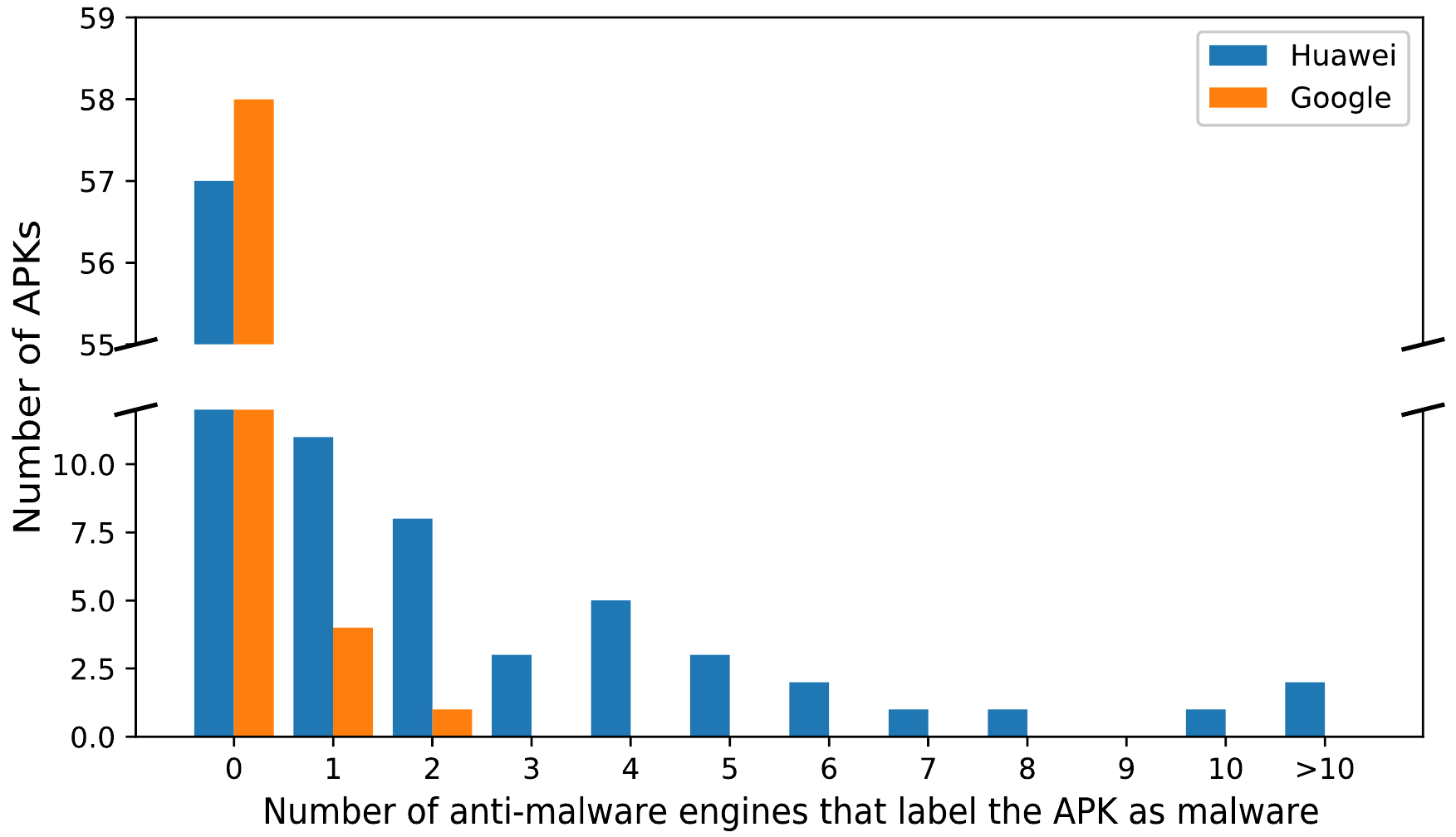}
  \caption{The performance of the VirusTotal in vetting fraudulent apps identified by \sysname.}
  \label{fig:virustotalres20210111}
\end{figure}

To compare the performance between \sysname and existing click fraud detection schemes, we use VirusTotal~\cite{vt}, a detection platform that integrates 60 anti-malware engines including Kaspersky~\cite{kaspersky} and McAfee~\cite{mcafee}, to double check the ad fraud apps identified by \sysname. Note that, as mentioned in~\cite{virustotalmeasurement, OpeningtheBlackboxofVirusTotal}, the performance of VirusTotal is not stable due to its design flaw. Therefore, to eliminate the detection error of VirusTotal, we uploaded those apps in July 2020 and January 2021 respectively to ensure the accuracy of the detection results, and the results are the same for both app sets. Although VirusTotal is not tailored for identifying humanoid attacks, we use it as a baseline since it provides basic functionality to spot some kinds of ad frauds as ``adware''. 
The results of VirusTotal and \sysname are shown in Fig.~\ref{fig:virustotalres20210111}. It is observed that 58 and 57 apps in Google Play and Huawei AppGallery successfully bypass all detection engines of VirusTotal, and only 5 apps can be detected by more than 7 engines. These results demonstrate that our \sysname outperforms existing detection engines in terms of click fraud detection.

\subsection{Humanoid Attacks in SDKs}
By conducting a longitudinal study on the aforementioned detected apps on Google Play.\footnote{Since Huawei AppGallery does not provide download channels for historical versions, we only studied those apps on Google Play, which were downloaded from apkpure~\cite{apkpure}.} We note that fraudulent SDK injection has played an increasingly important role in \attacks. We will elaborate upon the analysis below.

\subsubsection{Fraudulent SDKs in Humanoid Attacks}

\begin{figure}
  \centering
  \includegraphics[width=1\linewidth]{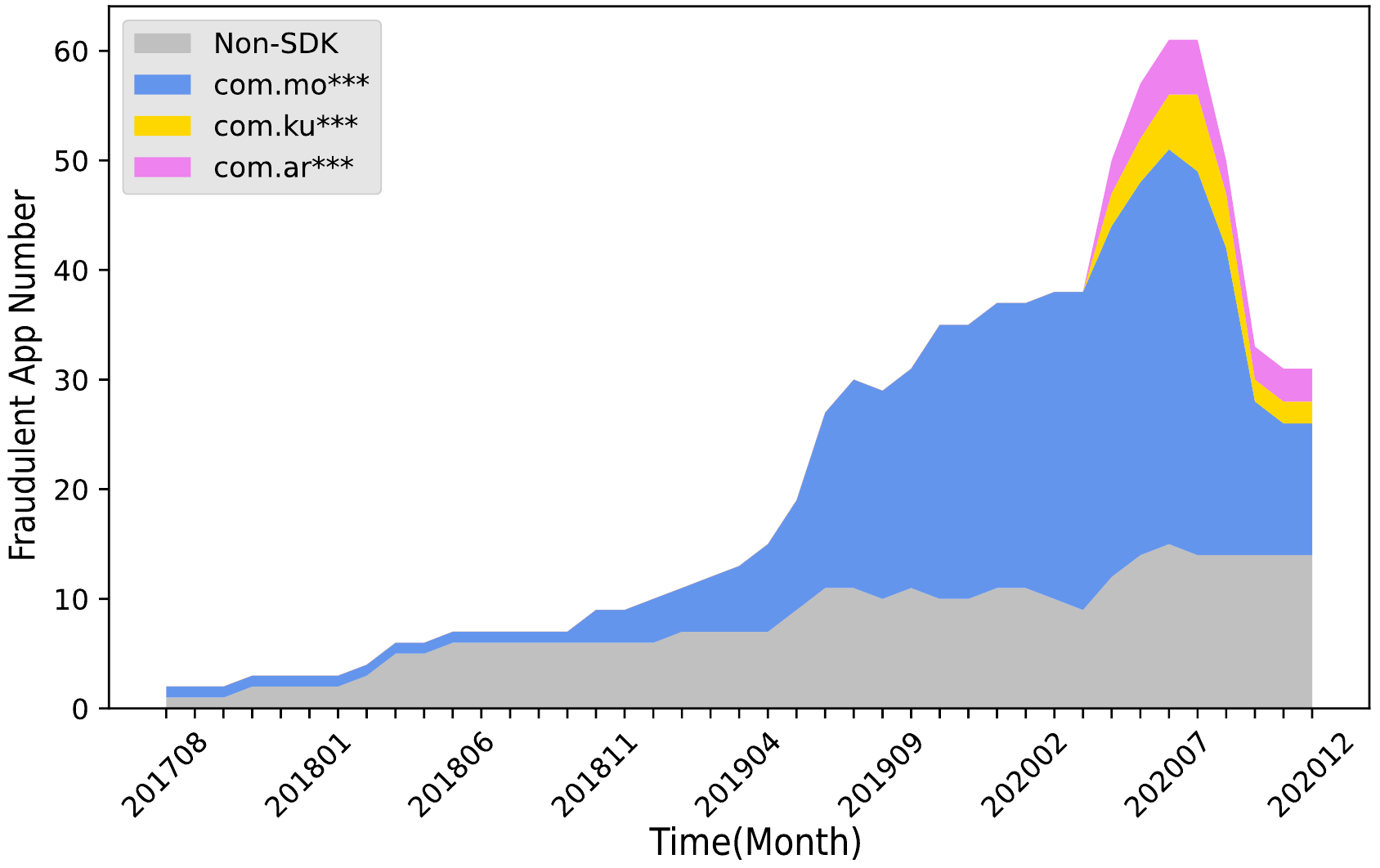}
  \caption{The stacked line chart of the number of fraudulent apps categorized by fraudulent SDKs from Google Play. }
  \label{fig:sdkratio}
\end{figure}

For the 157 apps detected by \sysname in \textsc{basic dataset}, we manually analyze the reasons leading to \attacks. It is observed that the \attack can be divided into non-SDK-based cases and SDK-based cases. In the former, the attackers directly inject the fraudulent click codes into the apps' local codes. However, in the latter case, the fake clicks are generated by the deployed third-party ad SDKs. As shown in Table~\ref{tab:SDKinAllDetectedRatio}, 67\% and 95.2\% fraudulent apps from Huawei AppGallery and Google Play belong to SDK-based cases, meaning that SDK based attack is the dominant manner of \attacks in our dataset.

For the 63 fraudulent apps in Google Play, we collect all 
their versions published from August 2017 to December 2020 and determine whether the attack is caused by the SDK. As shown in Fig.~\ref{fig:sdkratio}, it is observed that the proportion of ad SDK-based attack (\ie, injecting fraudulent codes into the third-party ad SDKs) has increased from 14\% in June 2018 to 83\% in August 2020, which means the SDK based attack is now a dominant attack approach. Besides, the most popular fraudulent SDKs in Google Play are com.mo***, com.ku***, and com.ar***. We observe the open source code on the Github, analyze the advertising SDK downloaded from official channels and confirm that all 7 SDKs labeled by ClickScanner were created fraudulently by SDK developers before publishing. The remaining 4 SDKs are not publicly available.

\subsubsection{New Findings on SDK com.mo***}

Another interesting finding from Fig.~\ref{fig:sdkratio} is that the attacks caused by SDK com.mo*** suffer an unusual decrease after August 2020.\footnote{We give more analysis of com.mo*** in Section \ref{subsec:monet}.} We analyze the apps deploying this SDK and find that 28 of the 39 apps have removed the SDK after August 2020. Then, by searching the commit record of it on GitHub, we also find that the \attack code was removed after the version published in November 2020. Inspired by a report from Forbes News in August 2020 revealing a malicious ad SDK named Mintegral~\cite{ming}, 
which was used by 1,200+ apps and was caught for performing click frauds since July 2019, we conjecture that the app developers may have noticed this issue and subsequently removed the ad SDK from their apps, or Google enforced a new policy against this SDK to disallow apps using it to be published in the Google Play Store. 

We also analyze the distribution of fraudulent SDKs of all 157 apps according to their categories. We show the detailed results in Appendix ~\ref{appen:The Distribution of Fraudulent SDKs} due to the page limitation, which shows that the use of SDK to commit \attack is not an isolated case against a certain category of apps, but a common method deserving attention.

\subsection{Analysis on the \textsc{Extensive Dataset}}\label{extensive analysis}
\textbf{Extensive dataset.} The aforementioned analysis is all based on the \textsc{basic dataset}. To present a comprehensive overview of \attacks, we continue to randomly collect 100,000 apps from the dataset published in~\cite{androzoo}, downloaded from Google Play updated on January 1, 2021, which are not in the \textsc{basic dataset}.\footnote{We did not collect apps from Huawei AppGalley since there are not as many apks available as on Google Play.}

\noindent \textbf{Measurement findings.} 
From 100,000 randomly selected apps, \sysname identifies 584 apps as fraudulent. 
Given that there is no ground truth for the identification of click fraud in our \textsc{basic dataset}, each app must be inspected to confirm the existence of click fraud. However, since inspecting
more than 500 apps is time-consuming, we instead choose to randomly sample 60 apps (>10\%) from the \textsc{extensive dataset}. After inspecting each app and identifying the click fraud actions, we show 100\% precision over the randomly sampled dataset and this suggests the effectiveness of \sysname. With more time and effort, manual check of a sample size greater than 200 apps (>30\%) would give a more pronounced precision rate. In view of the current best effort in manual verification, we highlight that currently there is no benchmark dataset publicly available for any accuracy-comparison of \attack identification approaches. Table~\ref{tab:SDKinAllDetectedRatio} shows that the proportion of SDK based cases among the \textsc{extensive dataset} is only 15.3\% which is much less than that in the \textsc{basic dataset} (\ie, 83.4\%). The possible reasons are as follows: 1) fraudulent SDKs are more likely to infect popular apps to commit more \attacks; 2) developers of popular apps pay more attention to app security compliance, and generally will not add fraudulent codes to apps themselves. Therefore, we reveal that the \attack has different properties depending upon app popularity, which will require different approaches for vetting.

\begin{table}[t]
\centering
\caption{Fraudulent app distribution by SDKs on Google Play and Huawei AppGallery}
\label{tab:SDKinAllDetectedRatio}
\setlength{\tabcolsep}{1.2mm}{
\begin{tabular}{@{}llcllll@{}}
\toprule
\multicolumn{2}{c}{\multirow{2}{*}{App market}}              & \multicolumn{2}{c}{\multirow{2}{*}{Total downloads}} & \multicolumn{3}{c}{Detection result} \\ \cmidrule(l){5-7} 
\multicolumn{2}{c}{}                                   & \multicolumn{2}{c}{}                               & SDK     & \# of apps    & Ratio     \\ \midrule
\multicolumn{2}{l}{\multirow{2}{*}{Huawei AppGallery}} & \multicolumn{2}{c}{\multirow{2}{*}{10,000}}        & yes        & 72          & 67.0\%    \\\cmidrule(r){5-7}
\multicolumn{2}{l}{}                                   & \multicolumn{2}{c}{}                               & no    & 22          & 33.0\%    \\\midrule
\multicolumn{2}{l}{\multirow{2}{*}{Google Play}}       & \multicolumn{2}{c}{\multirow{2}{*}{10,000}}        & yes        & 59          & 95.2\%    \\\cmidrule(r){5-7}
\multicolumn{2}{l}{}                                   & \multicolumn{2}{c}{}                               & no    & 4          & 4.8\%     \\\midrule
\multicolumn{2}{l}{\multirow{2}{*}{Google Play}}       & \multicolumn{2}{c}{\multirow{2}{*}{100,000}}       & yes        & 90          & 15.3\%    \\\cmidrule(r){5-7}
\multicolumn{2}{l}{}                                   & \multicolumn{2}{c}{}                               & no    & 494          & 84.7\%    \\ \bottomrule
\end{tabular}}
\end{table}

\begin{table*}
\centering
\caption{Runtime overheads evaluation of \sysname with other tools.}
\resizebox{0.99\linewidth}{!}{
\begin{tabular}{c|l|c|c|c|c|cccccc}
\hline
\multirow{2}{*}{Tools} & \multirow{2}{*}{FraudDetective~\cite{ndss2021}} & \multirow{2}{*}{FraudDroid~\cite{fengdongfrauddroid}} & \multirow{2}{*}{MAdFraud~\cite{crussellMadfraud}} & \multirow{2}{*}{DECAF~\cite{binliuDECFA}} & \multirow{2}{*}{AdSherlock~\cite{caoadsherlock}} & \multicolumn{6}{c}{\sysname}                         \\ \cline{7-12} 
                       &                                 &                             &                           &                        &                             & 0-20(MB) & 20-50(MB) & 50-100(MB) & 100-200(MB) & 200+(MB) & avg   \\ \hline
avg\_time(s)           & \multicolumn{1}{c|}{300}        & 216                         & 120                       & 675                    & 600                         & 16.37    & 20.82     & 16.11      & 16.79       & 22.02    & 18.42 \\ \hline
\end{tabular}}
\label{tab:Runtime Performance}
\end{table*}

\subsection{Time Overhead}
To evaluate the time overhead of \sysname, we first divide 120,000 apps from both the \textsc{basic} and \textsc{extensive} datasets into five categories with the apk sizes of $0 \sim 10$ MB, $10 \sim 50$ MB, $50 \sim 100$ MB, $100 \sim 200$ MB and above 200 MB respectively. Then we record the average running time of \sysname on each category. Note that the timeout of \sysname is set to 300 seconds, and only 742 out of 120,000 apps (0.62\%) do not terminate within 300 seconds.

We further compare the time overhead of \sysname with other previous tools (\eg, FraudDetective~\cite{ndss2021}, FraudDroid~\cite{fengdongfrauddroid}, MAdFraud~\cite{crussellMadfraud}, DECAF~\cite{binliuDECFA}, AdSherlock~\cite{caoadsherlock}, and  Clicktok~\cite{nagarajaclicktok}). 
Table~\ref{tab:Runtime Performance} lists the average time overhead of \sysname and other tools.

FraudDetective tests each app for five minutes. FraudDroid takes at most 216 seconds to analyze an app. MAdFraud needs 120 seconds to run on average. The mean and median time for analyzing an app
using DECAF is 11.8 minutes and 11.25 minutes respectively. AdSherlock executes each app in one emulator instance twice, which costs 10 minutes to extract ad fraud traffic patterns offline. Clicktok does not list its runtime performance but it also needs to execute each app for a chosen duration of time to interact with apps using the Monkeyrunner tool~\cite{Monkeyrunner}. It is observed that the average \sysname time cost for detecting \attacks is about 18.4 seconds per app. 

It is worth noting that even in the best case (\ie, 120 seconds in MAdFraud), the time overhead 
is nearly 6.5 times as much as \sysname. These results demonstrate that \sysname significantly outperforms existing tools in terms of the time overhead and that it is practical to deploy \sysname in the real world. Additionally, it is observed that when changing apk size, the detection time correlated to apk size is relatively stable.

\section{Case Study} \label{sec:case study}

In this section, we closely analyze the fraud apps discovered by \sysname. We present four representative \attack cases and explain newly obtained insights into how attackers commit mobile click fraud.

\subsection{Case 1. Humanoid Attacks after User's Legitimate Actions}

\begin{figure}
  \centering
  \includegraphics[width=\linewidth]{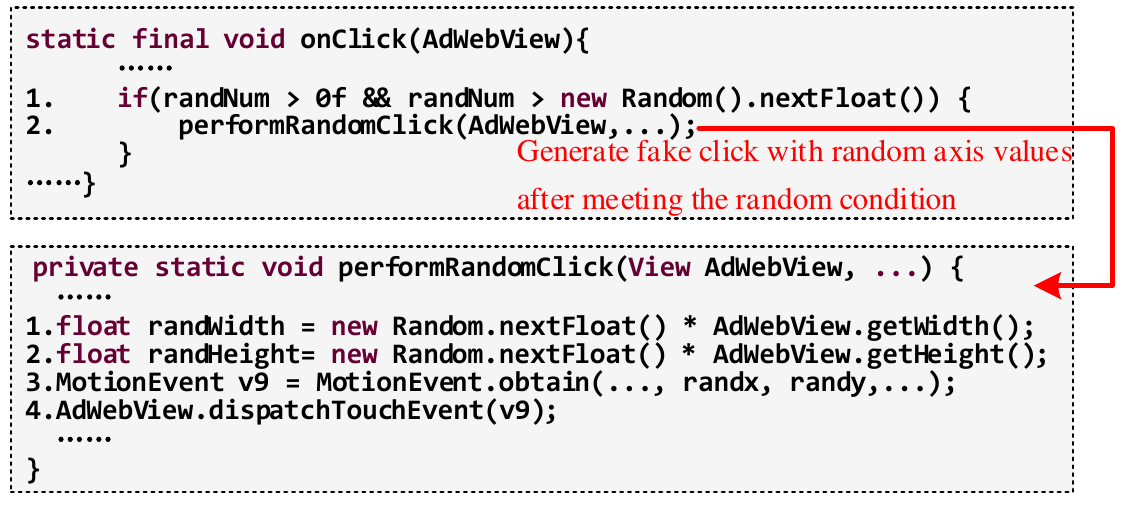}
  \caption{The code snippet from case 1 in the case study. The timing patterns of click fraud vary according to user actions.}
  \label{fig:casestudycode_cootek20210114}
\end{figure}

To fool the traditional detectors, some \attacks will follow the user’s legitimate actions. For instance, one fraudulent app, named com.co****.***********, is a communication and social app with a total of over 570 million downloads across all app markets searched on~\cite{kuchuan} and \sysname successfully reveals its process of generating \attacks in Fig.~\ref{fig:casestudycode_cootek20210114}. It first displays the advertisement to the users, and when the users click on the advertisement, it generates a set of random numbers in the trigger condition of fake clicks so that the \attack is triggered randomly. The same method is also used in the method of randomly generating click coordinates for \attacks. The potential threats with this app are as follows. $(i)$ Click frauds as such are difficult to detect with traditional detection methods, because their traffic patterns and click patterns vary with different users and are highly similar to real human clicks. Therefore, the traditional method of identifying click fraud that relies on the difference between patterns of click fraud and normal clicks is ineffective. $(ii)$ Attackers will even defraud advertisers for more advertising costs on the grounds that a considerable part of their users are interested in the advertisements they distribute and click twice. $(iii)$ We also find that the app is not alone in implementing fraudulent behaviors. Four other apps produced by the same company are manifested to have similar click fraud codes. Their total downloads reached over 658M across all app markets, therefore they have caused huge losses for advertisers.

\subsection{Case 2. Humanoid Attack that Can Adaptively Avoid Detection}\label{subsec:mtxx}

\begin{figure}
  \centering
  \includegraphics[width=1\linewidth]{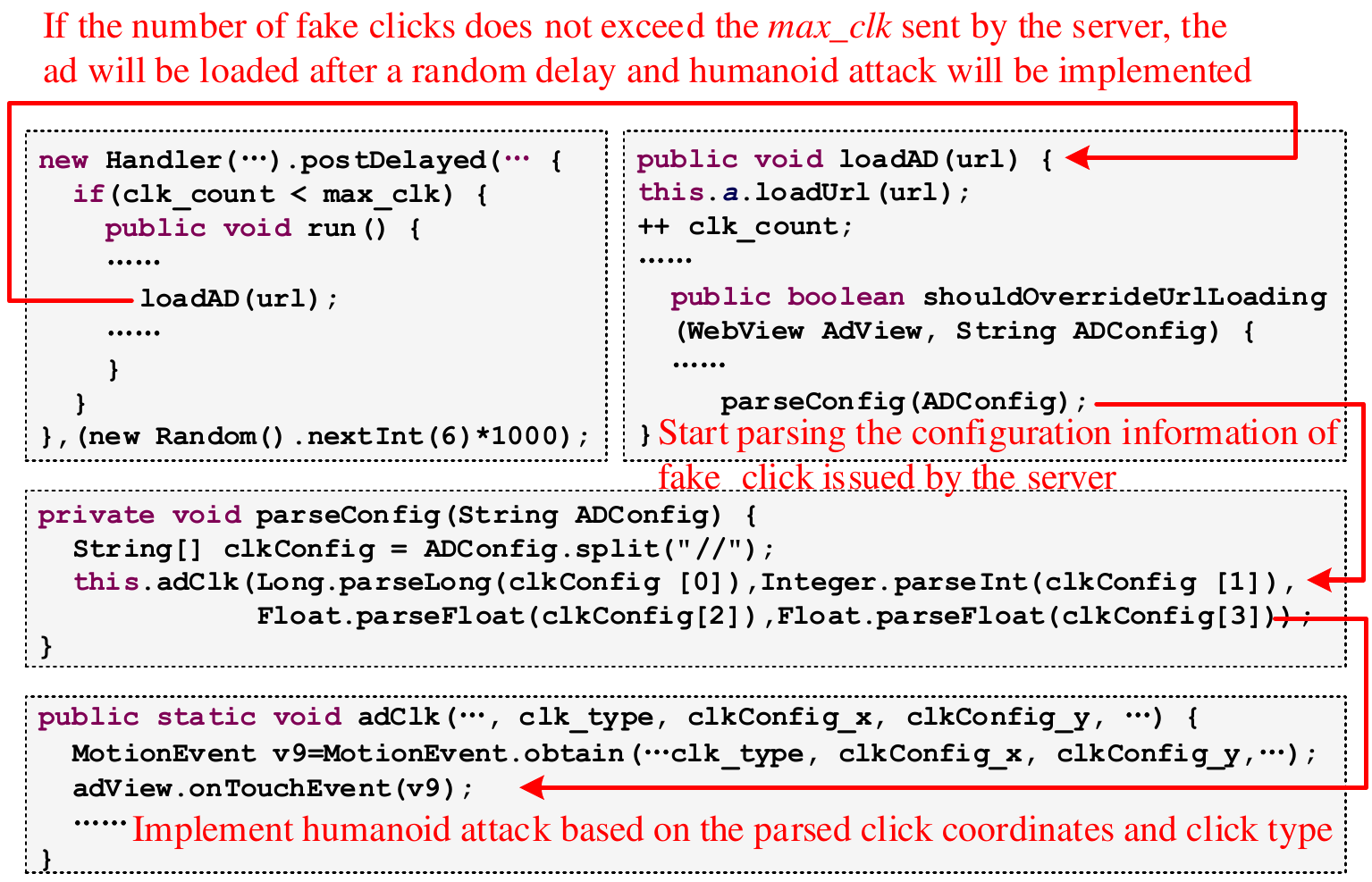}
  \caption{The code snippet from the case 2 in the case study. The configuration and commands of \attack come from the server, and the occurrence of fake click will be adaptively controlled locally to avoid detection.}
  \label{fig:casestudycode_mtxx20210115}
\end{figure}

Some apps predefine fake click's execution logic in their code and receive the click's configuration from remote servers, which can be changed at any time according to the situation. The occurrence of fake clicks will be adaptively controlled locally to avoid detection. As shown in Fig.~\ref{fig:casestudycode_mtxx20210115}, com.m*.****.**** is a photography app with over 9.5 billion downloads in all app markets searched on~\cite{kuchuan} and over 50 million installations since the time it was made available in Google Play. \sysname has uncovered it conducting \attacks against ad agencies. It first loads a configuration URL starting with "https://l**.u****.com:8***/l**/s?" and parses out data such as "ctr", "cvr", "max\_clk" and "max\_imp" in the returned JSON data structure as fake click configuration. Then it will count every time a fake click is triggered. When the number of fake clicks is greater than the value of the maximum number of clicks (max\_clk) sent by the remote server, it will stop conducting the \attack. If the number of fake clicks is smaller than "max\_clk", it will get the click properties, such as duration, action (up or down), and coordinates from the JSON data structure returned from the configuration URL to construct the \textit{MotinoEvent} object and implement the \attack. In doing so, the app can automatically execute a fake click on a random point in the ad view without the user actually clicking on the phone screen. The traffic of this fake click is identical to what would be generated by a real person, and hence the app adaptively avoids detection.

\subsection{Case 3. Humanoid Attack through Infected Ad SDKs}
\label{subsec:monet}

The above two apps involve only one developer (although the \attack code snippets involved in case 1 exist in multiple apps developed by the same company). As we mentioned in the Section~\ref{sec:evaluaiton}, fraudulent developers now have shifted from directly developing app applications to developing SDKs, which can make their click fraud code affect more apps. "com.mo***" is an advertising SDK with \attack codes involved in 43 apps out of 120,000 apps. It will override the \texttt{onClick()} method when initializing the ad view. This ad view will automatically click itself again when it is clicked by a user, which causes the ad view to be clicked twice in the advertiser’s view. This fraudulent SDK has a greater impact than the previous two because all apps that install this SDK will participate in fraudulent activities intentionally or unintentionally. The total number of app installations affected by this SDK reaches 270 million since they were made available on Google Play. It is worth noting that we found the source code of this SDK on the GitHub, and the developer deleted this code by himself on November 5, 2020. We guess this may be related to Google’s increasingly strict anti-ad-fraud measures.

\subsection{Case 4. Humanoid Attack by Disguising as an Auto-play Video Assist Function}

Some fraudulent activities are not only limited to clicking on static ads, but also pretend to be an accessibility service to help users to automatically play videos at the code level, but its real purpose is to click on the ad video. "com.iB***" SDK is a typical representative of this kind of \attack. It initializes a VideoWebView class in the VideoAdActivity function, and predefines an \texttt{onReadyPlay} callback function in this class. In this callback function, it will check whether the ad video is playing, and if it is not playing, it will automatically click the ad video. The "com.iB***" SDK forces people who want to leave because they are not interested in the ad video to watch the video to increase their ad revenue. The download number of the apps affected by this SDK is over 476 million across all app markets searched on~\cite{kuchuan}.
\section{Discussion} \label{sec:discussion}

\noindent \textbf{Scalability of \sysname.} 
In this work, we proposed a new time-saving and efficient framework for click fraud detection. To foster a broader impact, we implemented an efficient backward program slicing framework in \sysname which can detect the assignment process of any variable or parameters of any API. Hence, our framework can be easily extended by researchers for targeted and efficient security vetting of modern Android apps.\footnote{Publicly available at \url{https://github.com/Firework471/ClickScanner}.} In the future, we will also enhance \sysname to search over native code, and also extend it to other problems.

\noindent \textbf{Analysis of misclassification cases.} 
By manual analysis, the reasons for misclassifications are as follows:
For the 11 apps in the \textsc{seed apps} that are misclassified by \sysname: 1) False positives: some apps simulate a human click to obtain the focus of the window. Although Android provides a standardized API to achieve this function, obtaining focus through this inappropriate method is rare. 2) False negatives: some views, although they include ad contents, are not regarded as ad views by \sysname. 
For the 9 false positives in the \textsc{basic dataset}: 4 of them simulate a human click to obtain the focus of the window; 3 of them simulate a human click to automatically play a non-ad video; 2 of them are game apps, which achieve the expected game effect by shielding the original user's click event in the game webview and generating new click events.
To solve these issues, expending more effort on checking whether the click event target is the input box or non-ad video player in the webview, or improving the accuracy of ad view detection are both feasible solutions.

\noindent \textbf{Limitation 1: Fraud codes and instructions issued by remote servers.} One of the major limitations in \sysname is the inability to detect complex JavaScript or encrypted instructions that do not reside in the original fraudulent app and come from a command and control server, which is almost impossible to detect at the code level. To put a detail on \sysname, it can detect click fraud generated by the local code. However, if all the execution codes or commands are sent by a server at run time, due to the intrinsic problem of static analysis, \sysname can only know that the app will dynamically execute some code at a certain moment. Because \sysname cannot fetch the content of the code, it is impossible to know whether this code is used to perform click fraud. Hence, one future direction of this research is to support context-based static analysis to infer the purpose of the loaded code or instruction and analyze the dynamically loaded code using tools like DyDroid~\cite{DyDroid}.

\noindent \textbf{Limitation 2: Other types of ad views.} Although we have used one of the most popular and obfuscation-resilient tools, LibRadar~\cite{LibRadar}, with several constraints to identify ad views, our taxonomy may still be incomplete since it was built based on the current policies and literature available. This is also one of the main causes of the false-positives and false-negatives. Nevertheless, we can adjust the threshold appropriately to make a trade-off in the result, or we can use more tools~\cite{ReliableThird-PartyLibraryDetection, LibD} to identify ad libraries and ad views.
Additionally, due to the scalability of \sysname, \sysname is generic and can be extended to support the detection of click frauds on potential new types of ad views.

\noindent \textbf{Limitation 3: Obfuscation.} \sysname’s ability is constrained to its employed static analysis tool FlowDroid. \sysname can live well with lightweight obfuscation mechanisms since the logic of humanoid attacks is exposed. However, it cannot handle apps that adopt advanced obfuscations or packing techniques to prevent analysis of the app's bytecode. To address this issue, \sysname could actively interact with other techniques (\eg,  deobfuscation, unpacking, binary analysis) to recover the protected bytecode.

\section{Related Work} \label{sec:related}

In recent years, research in click fraud detection has mainly focused on dynamic analysis. Some of these approaches analyze ad network traffic and summarize a pattern of click-fraud traffic. Others rely on installing an additional patch or SDK on users' devices to check whether the ad click is fraudulent or not by checking the click pattern of touch events.

\noindent \textbf{Traffic patterns associated with ad network traffic.} Some previous works claimed that fraudulent clicks have different traffic patterns from benign ones. FraudDroid~\cite{fengdongfrauddroid} builds UI state transition graphs and collects their associated runtime network traffics, which are then leveraged to check against a set of heuristic-based rules for identifying ad fraudulent behaviors. MAdFraud~\cite{crussellMadfraud} automatically runs apps simultaneously in emulators in the foreground and background for 60 seconds each and found ad click traffic that occurred under the testing environment involving no user interaction. Clicktok~\cite{nagarajaclicktok} detects clickspam by searching for reflections of click traffic, encountered by an ad network in the past. Detection with traffic analysis depends primarily on the network traffic set gathered.

\noindent \textbf{Local pattern associated with click events.} Some previous works claimed that fraudulent clicks have different performance on users' devices compared to benign ones. FraudDetective~\cite{ndss2021} computes a full stack trace from an observed ad fraud activity to a user event and generates the causal
relationships between user inputs and the observed fraudulent activity. AdSherlock~\cite{caoadsherlock} injects the online detector into the app executable archive and marks the touch events as click fraud when the Android kernel does not generate a \textit{MotionEvent} object or the properties of the generated \textit{MotionEvent} object remain unchanged. DECAF~\cite{binliuDECFA} performs dynamic checking in an emulator and marks ad frauds if the layout or page context violates a particular rule. ClickGuard~\cite{clickguard} takes advantage of motion sensor signals from mobile devices since the pattern of motion signals is different under real click events and fraud events. However, ClickGuard will likely cause participants concern over data collection~\cite{perceptionsofonlinebehavioraladvertising,WhyJohnny,Unsafeexposure}.

These tools above played an important role in revealing the occurrence of mobile click fraud. However, if fraudulent apps (\eg, the apps implementing \attacks proposed in this study) simulate the real human's clicks patterns to bridge the gap between normal clicks and fake clicks in the click patterns, or hide their fraudulent behaviors while executing, or only trigger clicks in a certain period, models may fail at the very first stage. 
Crucially, most of the previous strategies cannot pinpoint which app class conducts click fraud.

\section{Conclusion} \label{sec:future and conclusion}

In this paper, we explored a new and sophisticated click fraud, named \attack, and we successfully revealed its attack patterns through static analysis. We designed and implemented \sysname for uncovering \attacks. 
By applying \sysname to measure real-world market apps containing 120,000 apps, \sysname identifies 157 fraudulent apps from 20,000 top-rated apps in Google Play and Huawei AppGallery. Our work also informs the impact of ad SDKs on click fraud and the distribution of fraudulent apps among different categories and popularity. In conclusion, our work suggests that the \attack is still widespread in the current app markets and we hope that the detection tool \sysname developed in this paper can effectively combat the emerging \attack.

\section*{Acknowledgement}
We thank the shepherd, Merve Sahin, and other anonymous reviewers for their insightful comments. We thank Jian Zhang and Zhushou Tang, affiliated to PWNZEN InfoTech Co., LTD, for their valuable assistance of our analysis of the motivating example. The authors affiliated with Shanghai Jiao Tong University were, in part, supported by the National Natural Science Foundation of China under Grant 61972453 and the National Natural Science Foundation of China under Grant 62132013. Minhui Xue was, in part, supported by the Australian Research Council (ARC) Discovery Project (DP210102670) and the Research Center for Cyber Security at Tel Aviv University established by the State of Israel, the Prime Minister's Office and Tel Aviv University. Xiaokuan Zhang was supported in part by the NortonLifeLock Research Group Graduate Fellowship.

\vfill\eject
\bibliographystyle{ACM-Reference-Format}
\bibliography{main}

\appendix
\section*{Appendix}

\section{Results of Fraudulent Apps and SDKs}\label{appen:Results of Fraudulent Apps and SDKs}

Table~\ref{tab:bigtable} lists some of the apps of a large download number which implement \attacks. The full list will be submitted to Huawei and Google. The download number is only from Google Play or Huawei AppGallery. The highest number of downloads (from Huaweu AppGallery) reached 3.7 billion. \sysname successfully locates where these click fraud occurs in codes, which will facilitate the digital forensics of these fraudulent behaviors. Meanwhile, we mark out whether these \attacks were caused by the advertising SDKs in the sixth column.

\begin{table*}[t]
\centering
\caption{Details of \attacks in apps from Google Play and Huawei AppGallery}
\label{tab:bigtable}
\scalebox{0.78}{
\begin{tabular}{@{}lllllc@{}}
\toprule
Package Name                                                & Version                  & Catgory                  & Download & Fraud Location                               & SDK based \\ \midrule
com.m*.****.****                                            & 8.7.1.5                  & Photography              & 3.7B     & com.wa***                                    & No        \\
com.ba***.********                                          & 7.33.0                   & Others                   & 200.0M   & com.ba***: void continueDispatchTouchEvent() & Yes       \\

com.if*****.*************                                   & 7.3.75                   & Music                    & 200.0M   & com.if***: void startClickConfirmBtn()
                      & Yes       \\
com.ut******.******                                         & 6.5.5                    & Video Players \& Editors & 100.0M   & com.mo***: void onClick()                    & Yes       \\
com.ac*********.*******                                     & 6.1.10              & Weather                  & 100.0M   & com.lo***: void e()                          & Yes       \\
com.co****.***********                                      & 6.8.0.4                  & Social                   & 100.0M   & com.co***: void dispatchTouchEvent()         & Yes       \\
com.me***.*******                                           & 3.9.6.0                  & Photography              & 100.0M   & com.ho***: void a()                          & No        \\
t*.yi***.****                                               & 5.10.2                   & Video                    & 100.0M   & com.an***: boolean a()                       & Yes       \\
com.co****.***********                                      & 6.6.4.2                  & Social                   & 100.0M   & com.co***: void dispatchTouchEvent()         & No        \\
com.ai**                                                    & 3.9.2.3076               & Education                & 96.11M   & com.dz***                                    & Yes       \\
com.ti****.********                                         & 4.3.8                    & Tools                    & 67.69M   & com.ti***: void BMa()                        & No        \\
com.qu*******.******                                        & 1.7.0.0                  & Tools                    & 57.22M   & com.iB***: void a()                          & Yes       \\
com.cl*******.********                                      & 1.4.7.2                  & Tools                    & 50.0M    & com.mo***: void onClick()                    & Yes       \\
com.pu***.*****.****                                        & 1.0.8                    & GAME                     & 50.0M    & com.ga***: void c()                          & Yes       \\
com.fu******.*****.******                                   & 3.5.8.7                  & Video                    & 46.97M   & cn.co***: void h()                           & No        \\
com.hu*****.*****************                               & 6.7.0.3                  & Others                   & 46.54M   & com.co***: void dispatchTouchEvent()         & No        \\
com.if*****.********                                        & 4.4.1264                 & Business                 & 42.73M   & com.if***: void run()                        & Yes       \\
com.st******.********                                       & 4.25                     & Lifestyle                & 41.11M   & com.xm***: void a()                          & No        \\
com.lw**.*******.**                                         & 1.29.32                  & Books                    & 33.9M    & com.lw***: void d()                          & No        \\
com.du******                                                & 4.44.2             & Education                & 31.04M   & e***: void a(java.lang.Object,java.util.Map) & No        \\
com.ma****.*******                                          & 2.4.0                    & Books                    & 30.64M   & com.ba***: boolean click()                   & No        \\
com.xm****.*******.***                                      & 2.31.5                   & Shopping                 & 29.15M   & com.by***: void a()                          & Yes       \\
com.dz.*****                                                & 3.9.2.3074               & Books                    & 24.92M   & com.dz***                                    & Yes       \\
com.iy*.******.**********                                   & 5.13.5.02                & Books                    & 24.73M   & com.iB***: void autoClick()                  & Yes       \\
com.xi******.********                                       & 3.9.2.3069               & Books                    & 20.81M   & com.dz***                                    & Yes       \\
com.hi.****.*****.******.**                                 & 1.7.2                    & Sports                   & 19.82M   & com.mo***: void c()                          & Yes       \\
com.xi******.*******                                        & 4.8.52                   & Video                    & 19.55M   & com.bi***: boolean O()                       & No        \\
com.ot*.*********                                           & 1.18                     & Music                    & 15.1M    & com.ot***: void run()                        & No        \\
com.pe*******.*****.**                                      & 3.1.0                    & Lifestyle                & 12.01M   & com.mo***: void c()                          & Yes       \\
com.is****                                                  & 3.9.2.3068               & Books                    & 10.71M   & com.dz***                                    & Yes       \\
com.ca*****.*******.****                                    & 3.4.7                    & Finance                  & 10.17M   & com.we***: void a()                          & No        \\
com.sc**********                                            & 6.11.0.3                 & Entertainment            & 10.0M    & com.mo***: void onClick()                    & Yes       \\
com.ta*******.*******                                       & 6.4.8                    & Communication            & 10.0M    & com.mo***: void onClick()                    & Yes       \\
com.li**.********.***                                       & 5.8.0                    & Communication            & 10.0M    & com.mo***: void onClick()                    & Yes       \\
com.im***.******                                            & 4.10.1.13493             & Video Players \& Editors & 10.0M    & com.mo***: void onClick()                    & Yes       \\
com.cl*******.***********                                   & 1.1.5.2                  & Tools                    & 10.0M    & com.ku***: void onViewClicked()              & Yes       \\
me*******.pr*.*********                                     & 1.3.7                    & Social                   & 10.0M    & com.mo***: void onClick()                    & Yes       \\
com.ba*******.****.*****.*****.***                & 9.0.10                   & Sports                   & 10.0M    & com.mo***: void onClick()                    & Yes       \\
com.cl*******.***********                                   & 1.1.5.2                  & Tools                    & 10.0M    & com.ku***: void onViewClicked()              & Yes       \\
com.sw*******.*****.****                                    & 2.5.3                    & Photography              & 10.0M    & com.ku***: void onViewClicked()              & Yes       \\
fm.ca*****.*********.*****.*******                          & 8.16.0         & Music \& Audio           & 10.0M    & d***: void onClick()                         & No        \\
com.po*****.********                                        & 6.4.0 & News \& Magazines        & 10.0M    & e***: void onClick()                         & No        \\
m*.go*********.****                                         & 8.1.5                    & Music \& Audio           & 10.0M    & com.mo***: void onClick()                    & Yes       \\
lo*****.fl*****.******.*******              & 1.2.9                    & Health \& Fitness        & 10.0M    & d***: void a()                               & No        \\
in******.he********.********* & 1.0.6                    & Health \& Fitness        & 10.0M    & Ql***: void a()                              & No        \\
com.aw*.*******                                             & 5.17.2-10                & Weather                  & 10.0M    & com.lo***: void d()                          & Yes       \\
ca****.**.*****.******.*****                                & 1.8.7                    & Communication            & 10.0M    & com.mo***: void onClick()                    & Yes       \\
com.xm****.***                                              & 1.9.4                    & Tools                    & 9.68M    & com.by***: void a()                          & Yes       \\
com.di*************                                         & 3.9.2.3074               & Others                   & 7.85M    & com.dz***                                    & No        \\
com.lb*.********                                            & 6.1.2562                 & Tools                    & 7.09M    & pp***: void a(android.view.View,float,float) & No        \\
com.if*****.*************.*********                         & 7.2.67                   & Music                    & 7.02M    & com.iB***: void autoClick()                  & Yes      \\

com.me*******.*********.*******                             & 5.8.0                    & Tools                    & 5.0M     & com.mo***: void onClick()                    & Yes       \\
com.am***.********.***.*******.*****.****                   & 4.7.0.693         & Others                   & 5.0M     & m***: void onClick()                         & No        \\
m*.ne******.***********.**********                          & 2.6.7                    & Communication            & 5.0M     & c***: void onClick()                         & No        \\
mo**.in******.*********.******.*******.********.***         & 16.6.0.50080             & Weather                  & 5.0M     & com.mo***: void onClick()                    & Yes       \\
com.wo**.******.***.*********.******                        & 1.1.7                    & GAME                     & 5.0M     & com.ga***: void c()                          & Yes       \\
                      com.fl***.****************                                  & 6.6.6.1                  & Others                   & 530.0K   & com.co***: void dispatchTouchEvent()                     & No \\
com.xi*****.*****                                           & 2.20.5                   & Shopping                 & 480.0K   & com.iB***: void a()                                      & Yes \\
com.me**********.****************                           & 6.0                      & Education                & 470.0K   & c***: void a(java.lang.Object,java.util.Map)             & Yes \\
com.yz*.*******.******                                      & 1.07                     & Others                   & 250.0K   & com.da***: MotionEvent createMotionEvent()  & No\\

com.zh************.***********                              & 2.3.8                    & Books                    & 250.0K   & com.zh***: void clickView()                              & No  \\
com.yz*.*******.******                                      & 1.07                     & Others                   & 250.0K   & com.da***: MotionEvent createMotionEvent()  & No  \\
com.ha*****.******                                          & 1.3.0                    & Travel                   & 230.0K   & com.iB***: void autoClick()                              & Yes\\
com.ha*****.******                                          & 1.3.0                    & Travel                   & 230.0K   & com.iB***: void autoClick()                              & Yes       \\

com.bu****.******                                           & 1.5.8                & Productivity             & 100.0K   & com.bu***: void emulateClick()                           & No \\
com.we******.******                                         & 2.2.4                    & Video                    & 20.0K    & com.we***: void simulateClick()                          & No        \\
com.se*****.****                                            & 2.3.10                   & Tools                    & 10.0K    & com.se***: void createClickEvent()                       & No        \\

\bottomrule
\end{tabular}}
\end{table*}

\section{The Distribution of Apps Affected by Humanoid Attacks}\label{appen:humanoid:category}
\begin{figure}
  \centering
  \includegraphics[width=\linewidth]{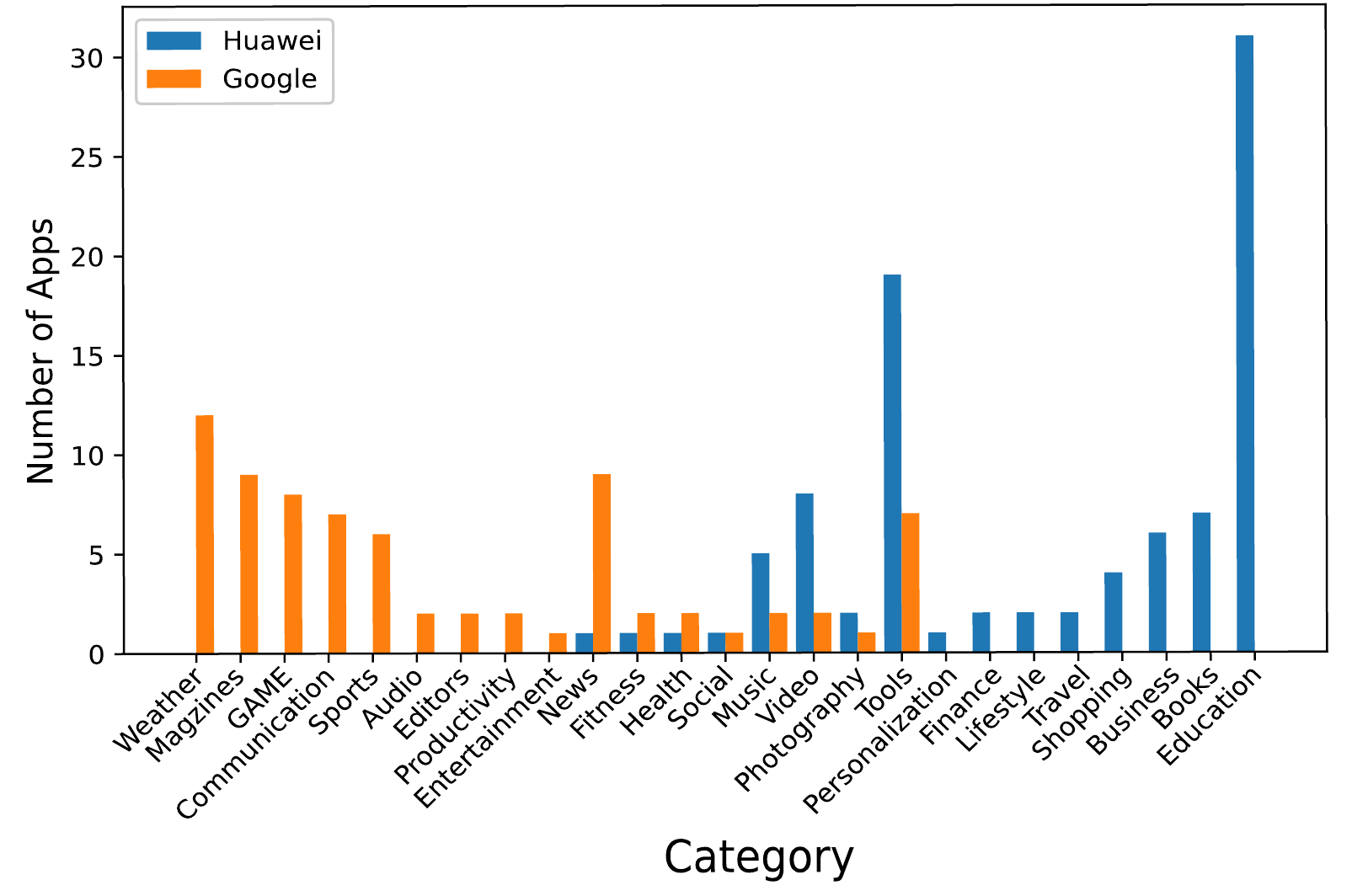}
  \caption{The number of fraudulent apps group by different categories on Google Play and Huawei AppGallery.}
  \label{fig:historical:apps}
\end{figure}

To quantify the damage of \attacks in the real world, it is important to know the category distributions of identified malicious apps. Thus, for each fraudulent app, we extract its category (\eg, books, education, weather) labeled by app markets and illustrate the statistical results in Fig.~\ref{fig:historical:apps}. It is observed there is a significant difference between the categories of fraudulent apps from Google Play and Huawei AppGallery. For instance, among 25 categories, fraudulent apps from both app markets exist in only 8 categories. 

Since the target users in Google Play and Huawei AppGallery are different (\ie, the former mainly targeting U.S. and Europe, and the latter mainly targeting China), the difference of marketing is perhaps caused by the interest/cultural differences of users, and the users' interests in mobile ads~\cite{198139}.
For instance, the fraudulent apps detected in Huawei AppGallery 
are concentrated in ``Education'', ``Books'' and ``Shopping'' because China has a prosperous online shopping industry and a cultural emphasis on education, while in Google Play, the detected apps are concentrated in ``News'', ``Magazines'' due to the diversity of media in society. 
It is also worth noting that there are apps from the ``Tools'' category marked by \sysname in both markets.
This may be because it is easier to obtain system permissions to hide fraudulent activities for these types of apps. In summary, these discoveries indicate that application markets in different regions need to focus on vetting different types of applications.

\section{The Distribution of Fraudulent SDKs}\label{appen:The Distribution of Fraudulent SDKs}

\begin{figure}
  \centering
  \includegraphics[width=1\linewidth]{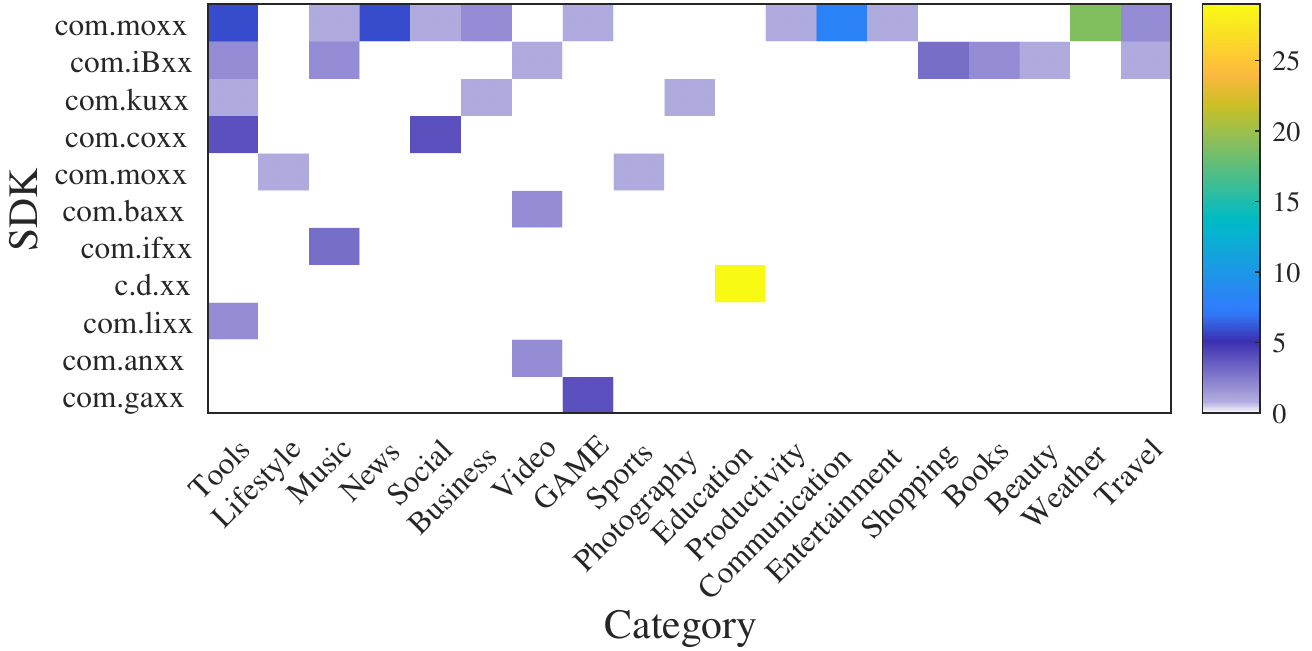}
  \caption{SDKs across app categories. The color encodes the number of apps involving fraudulent SDKs.}
  \label{fig:sdkviacatgory}
\end{figure}

We also analyze the distribution of fraudulent SDKs of all 157 apps according to their categories. 
As shown in Fig.~\ref{fig:sdkviacatgory}, the fraudulent apps existing in nearly half of the categories (19/45) are affected by 11 SDKs in total. This shows that the use of SDK to commit \attack is not 
an isolated case against a certain category of apps, but a common method deserving attention. Furthermore, it is observed that the com.mo*** SDK and com.iB*** SDK have infected a large proportion of applications. Therefore, to effectively thwart \attacks, it is vital to put efforts on vetting SDKs 
as well as the apps.

\end{document}